\documentclass[notitlepage,letterpaper,showpacs,preprintnumbers,amsmath,amssymb,nofootinbib,twocolumn, superscriptaddress]{revtex4-1}
\pdfoutput=1
\usepackage{amssymb,amsmath,latexsym,mathrsfs}
\usepackage{graphicx}
\graphicspath{{plots/}}
\usepackage{bm}
\usepackage[normalem]{ulem}
\usepackage[usenames,dvipsnames]{color}
\usepackage{comment}
\usepackage{lipsum}
\usepackage{hyperref}
\hypersetup{
    colorlinks=true,       
    }   
\usepackage{xspace}
\usepackage{soul,xcolor}

\newcommand{\cosmomc}{\texttt{CosmoMC}\xspace}
\newcommand{\camb}{\texttt{CAMB}\xspace}

\newcommand{\lcdm}{$\Lambda$CDM}
\newcommand\be{\begin{equation}}
\newcommand\ee{\end{equation}}
\newcommand\bea{\begin{eqnarray}}
\newcommand\eea{\end{eqnarray}}

\newcommand{\neff}{N_{\rm eff}}
\newcommand{\nsat}{N_{\rm sat}}

\newcommand\eq[1]{Eq.~(\ref{#1})}

\begin{document}

\setstcolor{red}

\title{Cosmological searches for a non-cold dark matter component}

\author{Stefano Gariazzo}
\affiliation{Instituto de F\'isica Corpuscular (IFIC),
CSIC-Universitat de Valencia,\\
Apartado de Correos 22085,  E-46071, Spain}

\author{Miguel Escudero}
\affiliation{Instituto de F\'isica Corpuscular (IFIC),
CSIC-Universitat de Valencia,\\ 
Apartado de Correos 22085,  E-46071, Spain}

\author{Roberta Diamanti}
\affiliation{GRAPPA Institute of Physics, University of Amsterdam,\\
Science Park 904,1098 GL Amsterdam, Netherlands}

\author{Olga Mena} 
\affiliation{Instituto de F\'isica Corpuscular (IFIC),
CSIC-Universitat de Valencia,\\ 
Apartado de Correos 22085,  E-46071, Spain}

\begin{abstract}
We explore an extended cosmological scenario where the dark matter is an admixture of cold
and additional non-cold species. The mass and temperature of the non-cold dark matter particles are
extracted from a number of cosmological measurements. Among others,
we consider tomographic weak lensing data and Milky Way dwarf
satellite galaxy counts.
We also study the potential of these scenarios in alleviating
the existing tensions between local measurements
and Cosmic Microwave Background (CMB) estimates of the
$S_8$ parameter, with $S_8=\sigma_8\sqrt{\Omega_m}$, and of the Hubble constant $H_0$. In principle, a sub-dominant, non-cold dark matter
particle with a mass $m_X\sim$~keV, could achieve the goals above. However,
the preferred ranges for its temperature and its mass 
are different when extracted from weak lensing observations and from Milky Way dwarf satellite galaxy counts,
since these two measurements
require suppressions of the matter power
spectrum at different scales. Therefore, solving
simultaneously the CMB-weak lensing tensions and the small scale crisis in the standard
cold dark matter picture via only one non-cold dark matter
component seems to be challenging.
\end{abstract}


\maketitle


\section{Introduction}
Within the current canonical cosmological model, dubbed the $\Lambda$CDM
model, the dark matter is assumed to be made of a totally cold gas of weakly
interacting particles, accounting for $\sim26\%$ of the current
universe mass-energy density. This standard picture 
has been extremely successful in explaining both the large scale
structure observations of our universe and the pattern of the
temperature and polarization fluctuations in the Cosmic Microwave
Background (CMB)~\cite{Ade:2015xua}.
Nevertheless, the mechanism explaining
the origin and the physics of this cold dark matter component remains
obscure~\cite{Bertone:2004pz, Bergstrom:2012fi, Kusenko:2013saa}, with
possible candidates ranging from the GeV-TeV energy scale to very
light ($\mu$eV) dark matter axions. Together with this hitherto theoretically unknown cold dark matter
nature, there are a number of observations which further motivate the
searches for other possible dark matter candidates.

On the one hand, there is the \emph{small scale crisis} of the $\Lambda$CDM
model.
This problem is closely related to several galactic and
sub-galactic phenomena, as the \emph{Milky Way satellites
problem}~\cite{Klypin:1999uc, Moore:1999nt} and the so-called
\emph{too big to fail problem}~\cite{BoylanKolchin:2011dk}, which
refer to the fact that
the predictions from the $\Lambda$CDM picture fail in reproducing the number of low-mass subhalos expected
within a Milky Way-sized halo and the measured kinematics of the Milky 
Way satellites, respectively. A large effort in the literature has
been devoted to alleviate these
problems~\cite{Wang:2016rio,Lovell:2016nkp,Sawala:2012cn,Sawala:2015cdf,Fattahi:2016nld,Polisensky:2013ppa,Vogelsberger:2012ku,Schewtschenko:2015rno,Lovell:2011rd,Lovell:2013ola,Lovell:2015psz,Nakama:2017ohe}.

On the other hand, recent measurements of tomographic weak
gravitational lensing, as those from the Kilo Degree-450 deg$^2$
Survey (KiDS-450)~\cite{Hildebrandt:2016iqg,Joudaki:2016kym}, show  substantial discordances
with CMB measurements from
Planck~\cite{Ade:2015xua,Aghanim:2015xee} in the matter perturbations at small scales.
These discordances are quantified in terms of the extracted values of
the amplitude of the small-scale density fluctuations,
quantified by the parameter $\sigma_8$, at a given matter density,
$\Omega_m$.
In particular, the quantity $S_8=\sigma_8\sqrt{\Omega_m}$
as measured by KiDS is in tension with the Planck estimate at the
level of $2.3\sigma$~\cite{Hildebrandt:2016iqg}. Similar results had already appeared in the past 
from the analyses of CFHTLenS
data~\cite{Kilbinger:2012qz,Heymans:2013fya}. A number of recent 
dedicated studies in the literature have shown that the CFHTLenS and the
KiDS discrepancies are independent of the small-angle
approximations commonly exploited in weak lensing data analyses~\cite{Lemos:2017arq,Kilbinger:2017lvu,Kitching:2016zkn}.

Here, instead of refining cosmic shear analyses, we follow a different
avenue to ameliorate these problems. In the spirit of Ref.~\cite{Joudaki:2016kym}, we consider a modified
version of the most economical pure cold dark matter model, allowing
for a mixed dark matter cosmology with an additional, non-totally
cold, dark matter relic. These models with an admixture of cold and non-cold dark
matter particles have been dubbed mixed dark matter (MDM) models; see
e.g.~Refs.~\cite{Palazzo:2007gz,Boyarsky:2008xj,Maccio:2012rjx,Anderhalden:2012qt,Anderhalden:2012jc,Diamanti:2017xfo}.
The motivation to consider
these models is twofold: in addition to their potential in alleviating the
tension between cosmic-shear and CMB measurements, they could also 
provide a solution to the aforementioned
$\Lambda$CDM small-scale crisis, while leaving unchanged the predictions from
the $\Lambda$CDM model at large scales. The reason is simple: the
particle associated to the second, non-totally cold dark matter
component will have a significant free-streaming length, 
that affects the matter power spectrum on the smallest scales,
therefore improving the compatibility with the observations of the
local Universe~\cite{Weinberg:2013aya} through a reduction of the $S_8$ quantity.

In this study we scrutinize these mixed dark matter scenarios, using the
most recent tomographic weak lensing measurements from the KiDS-450
survey, in combination with Planck CMB data. By means of these datasets we shall
derive constraints on the current temperature and mass-energy density of the
non-cold dark matter component, searching for the most favored cosmological dark matter scenario. Furthermore, we also consider
current estimates from the observed number of Milky Way satellite
galaxies, comparing these results to those preferred by weak
lensing data.

The structure of the paper is as follows. 
In Sec.~\ref{sec:method}
we present the methodology followed here, describing the Mixed Dark
Matter model, the included datasets and the technical details of our
numerical analyses. Our results are shown in
Sec.~\ref{sec:results}. We draw our main conclusions in Sec.~\ref{sec:conclusions}.

\section{Methodology}
\label{sec:method}
\subsection{Mixed Dark Matter modeling}
\label{ssec:dm}
In this paper we explore a scenario where the dark matter fluid is made of two
components: one which corresponds to
the standard cold dark matter (CDM) plus a second one with a non-zero
temperature $T_X$~\footnote{We consider relics with a Fermi-Dirac
  distribution. A change in the distribution will not change
  dramatically the results, see e.g.~\cite{Diamanti:2017xfo}.}.
In the following we shall refer to this
model as the Mixed Dark Matter (MDM) model.
While the CDM component is simply parameterized via its energy density
$\omega_c\equiv\Omega_c h^2$, the second dark matter
component is parameterized through its temperature $T_X$ and its energy density fraction $f_X$ relative to the
total dark matter component $\omega_{DM}=\omega_{X}+\omega_{c}$, defined as
\begin{equation}\label{eq:ratio}
f_X=\frac{\omega_X}{\omega_{DM}}\,,
\end{equation}
where $\omega_i\equiv\Omega_i h^2$ refers
to the present mass-energy density.
Notice that the mass of the second, non-cold dark matter particle can
be computed from its energy density and its temperature using
the relation
\begin{equation}
\label{eq:mwdm}
\omega_X=\left(\frac{T_X}{T_\nu}\right)^3\,\left(\frac{m_X}{94\text{ eV}}\right)\,.
\end{equation}
In our analyses we shall consider $0\le f_X\le 1$ for the WDM energy density
and $-1.5\le \log_{10}(T_X/T_\nu)\le 0$ for its temperature. The upper
temperature prior, $T_X = T_\nu$, is fixed to be the one corresponding
to the pure \emph{hot} dark matter regime, i.e.~to the standard neutrino temperature,
while the lower prior is
chosen in order to preserve the validity of the numerical calculations for
the MDM model used here, as we explain
in what follows.

A crucial point when dealing with the MDM modeling
is related to the power spectrum of the density perturbations, which
is modified when a second non-totally cold dark matter component is
also considered in the cosmological evolution.
The deviations of the matter power spectrum in the MDM model
from its standard shape within the CDM model may be highly non-trivial
and must be treated cautiously,
since we are dealing with weak
lensing probes, which require a good knowledge of
the perturbation behavior in the non-linear regime~\footnote{This is a delicate
issue, as we are dealing with a second dark matter component different
from the standard three neutrino active contribution. For the implementation of massive
neutrinos in non-linear matter power spectrum simulations, see
Ref.~\cite{AliHaimoud:2012vj}.}. We recall that the non-linear approximations that are commonly used
for the numerical computations are calibrated on N-body simulations. These calibrations, however, must be considered carefully,
since the extrapolation for unusual models may spoil the correctness
of the adopted formulae.

\begin{figure*}[tp]
\begin{tabular}{cc}
\includegraphics[width=0.48\textwidth]{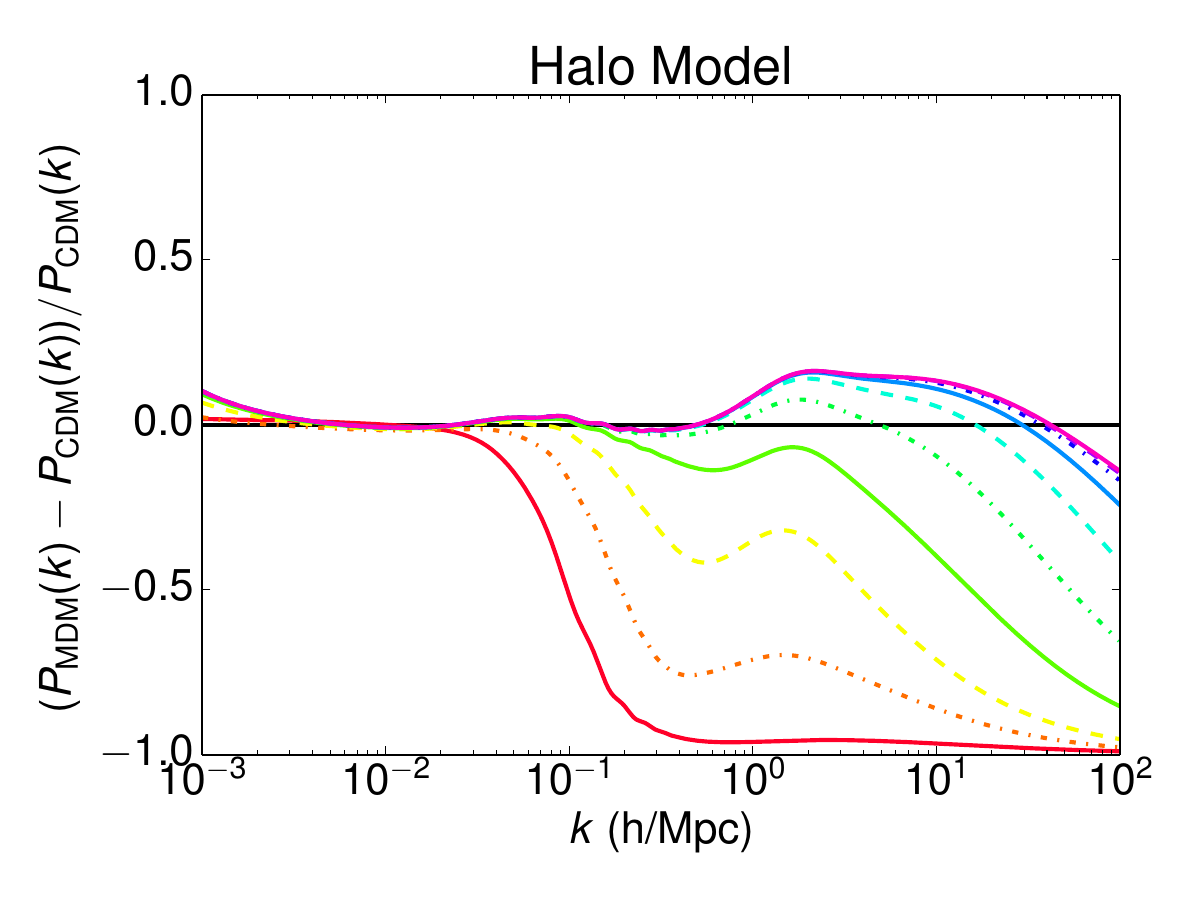} & \includegraphics[width=0.48\textwidth]{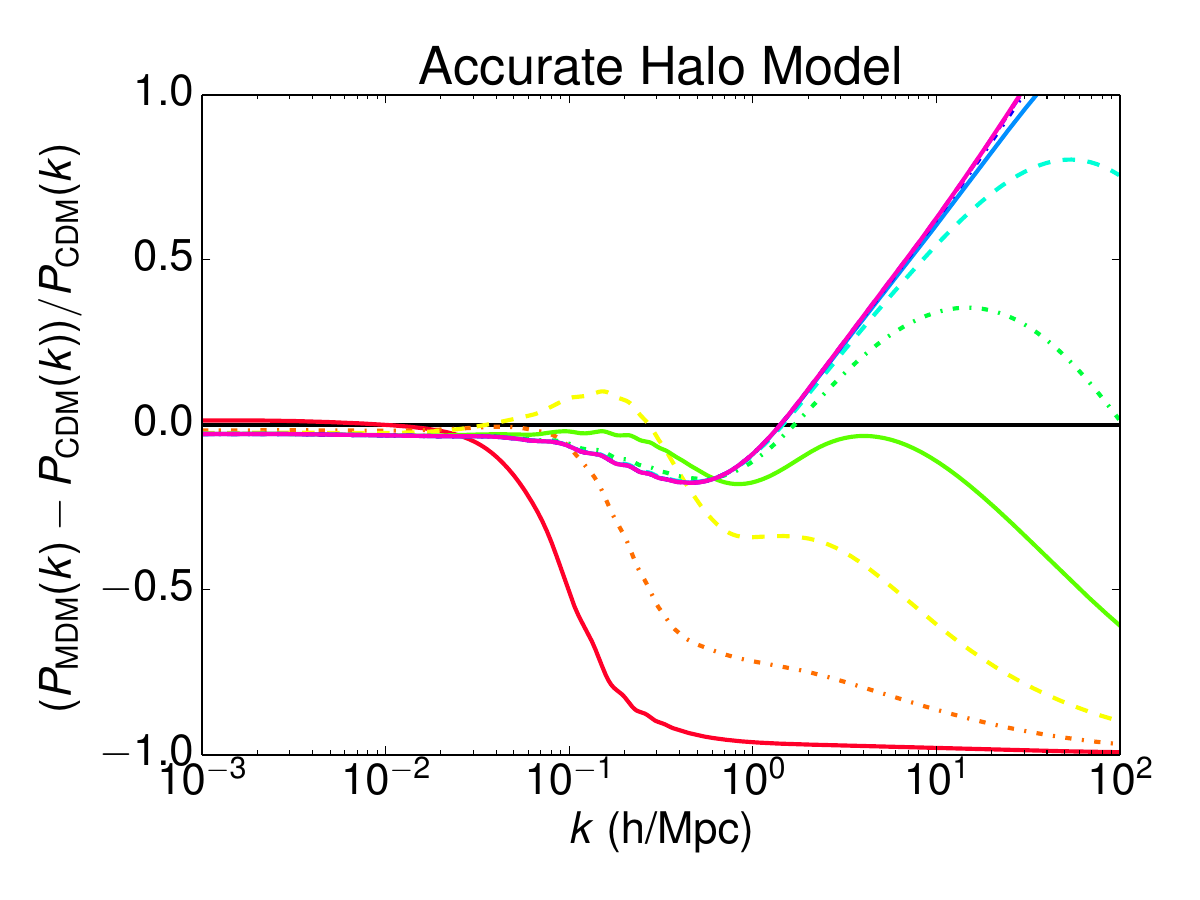} \\
\includegraphics[width=0.48\textwidth]{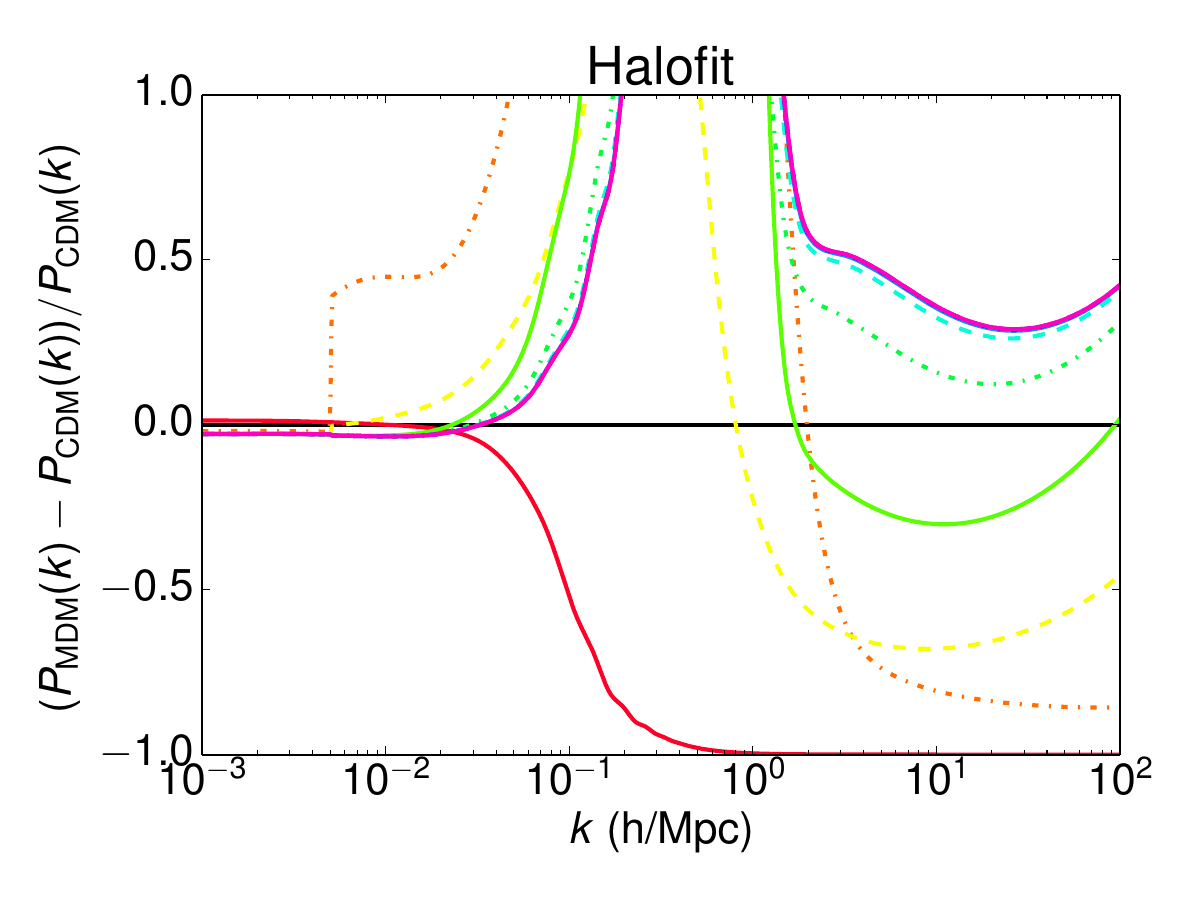} & \includegraphics[width=0.48\textwidth]{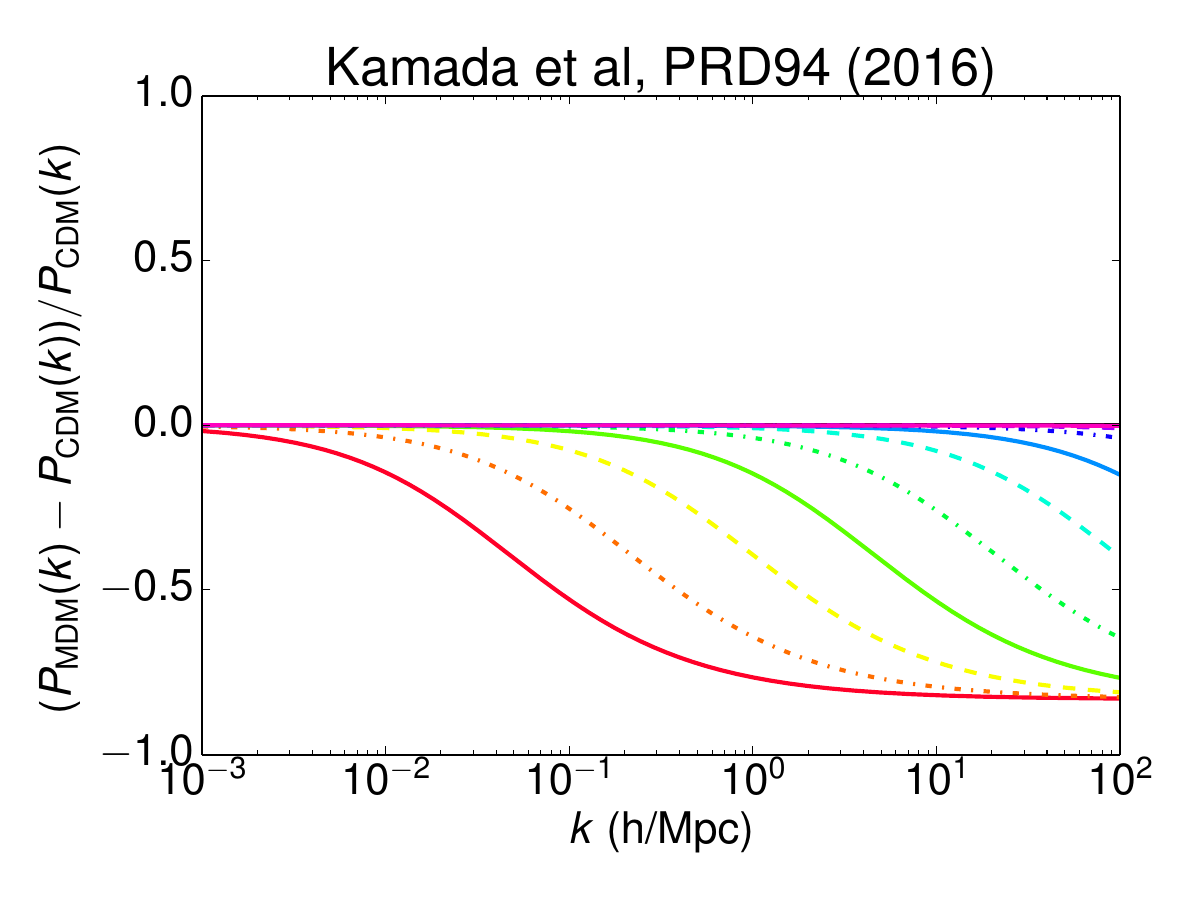}
\end{tabular}
\includegraphics[width=0.92\textwidth]{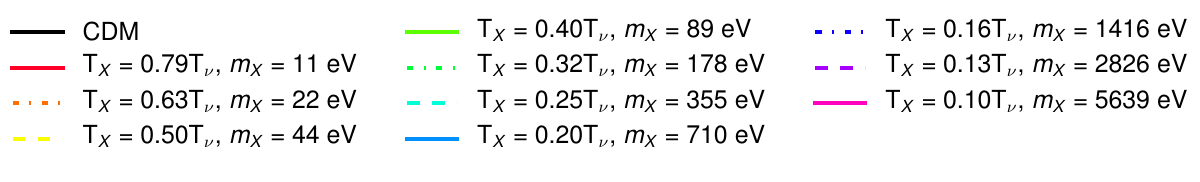}
\caption{\label{fig:nonlin}
Relative difference of the MDM non-linear matter power spectrum at $z = 0$ 
(for $f_X=0.5$) with different temperatures $T_X$, with respect to
the pure CDM case (for which $f_X=0$), for four different non-linear approaches, as described in the text.
For all of them we fix $\omega_{DM}=0.12$.
The non-linear prescription adopted in this work,
given by Ref.~\cite{Kamada:2016vsc},
is the one corresponding to the \textit{lower right} panel.}
\end{figure*}

In this regard, we show in 
Fig.~\ref{fig:nonlin} the relative difference between the
non-linear matter power spectrum of some of the MDM models explored
here with respect to the corresponding CDM only case.
We use $\omega_{DM}=0.12$ for all the plots, with $f_X=0$ for the CDM-only model
and $f_X=0.5$ for the other cases, and we vary the non-cold dark matter
particle temperature. The panels refer to four possible non-linear prescriptions:
the standard~\cite{Seljak:2000gq,Peacock:2000qk,Ma:2000ik,Cooray:2002dia} (upper left panel)
and the
accurate~\cite{Mead:2015yca} (upper right panel) versions of the halo model,
the standard \texttt{halofit} code~\cite{Takahashi:2012em}
(lower left panel) and the fitting formula presented in
Ref.~\cite{Kamada:2016vsc} (lower right panel),
that we shall adopt here (see below).
Notice that the \texttt{halofit} prescription badly
fails in reproducing the expected behavior of the non-linear power
spectrum when the temperature $T_X$ of the non-totally cold dark
matter particle deviates significantly from the neutrino temperature, $T_\nu$.
In the limit $T_X\rightarrow 0$, the MDM case should approach the CDM one,
eventually overlapping with it and this behavior is clearly not reproduced.
The accurate halo model~\footnote{The term ``accurate'' is related to
the fact that this improved version of
the original halo model~\cite{Seljak:2000gq,Peacock:2000qk,Ma:2000ik,Cooray:2002dia} takes into account several
corrections (that include, among others, the baryonic feedback),
i.e.\ factors that are not included in the standard halo model, see
Ref.~\cite{Mead:2015yca} for details.} also presents some problems at
the smallest scales. Following these results, the best model to
describe the non-linear perturbation growth in the MDM case seems to be
the standard halo model. However, even in this case there
exists an unphysical bump at scales $k\sim 1$~$h$/Mpc which makes its
predictions unreliable.

The clear failure of these three widely used non-linear models for a significant
range of temperatures $T_X$ has motivated us to
search for an alternative description of the non-linearities in the
power spectrum in the presence of an additional non-cold dark matter
component. We have therefore adopted the prescriptions presented
in Ref.~\cite{Kamada:2016vsc}, that we briefly summarize here.

Starting from the standard non-linear
matter power spectrum for a CDM universe $P_{\rm CDM}$
computed with \texttt{halofit}~\cite{Takahashi:2012em,Takahashi:2013sna},
the authors of Ref.~\cite{Kamada:2016vsc} find that the fitting function that best matches
the results of N-body simulations in the presence of a non-cold dark
matter component is given by:
\begin{eqnarray}
\frac{P_{\rm MDM}(k)}{P_{\rm CDM}(k)}&=&T^2(k; r_X, k'_d)\nonumber\\
&=&(1-r_X)+\frac{r_X}{(1+k/k'_d)^{0.7441}}
\label{eq:pmdm}
\,,
\end{eqnarray}
where the two quantities that appear in the right hand side read as
\begin{eqnarray}
\label{eq:fwarm}
r_X(f_X)&=& 1-\exp
\left(-a\,\frac{f_X^b}{1-f_X^c}\right)\,,
\\
\label{eq:kdprime}
k'_d(k_d,f_X)
&=&
k_d\cdot f_X^{-5/6}
\,.
\end{eqnarray}
In the latter equation, $k_d$ is the damping scale
given in Ref.~\cite{Inoue:2014jka} as a function
of the linear growth rate $D(z)$:
\be
k_d(m_X, z)
=
\left(\frac{m_X}{\text{keV}}\right)^{2.207}
D(z)^{1.583}
\,\,
388.8\,h\,\text{Mpc}^{-1}
\,.
\ee
Finally, parameters $a$, $b$, $c$ in \eq{eq:fwarm} are
obtained by fitting the parameterization above to the N-body simulation results~\cite{Kamada:2016vsc}:
\be
a=1.551\,,\quad b=0.5761\,,\quad c=1.263\,.
\ee
In all the relevant parts of our computation, therefore,
we substituted the non-linear matter power spectrum with the one given by \eq{eq:pmdm}.
This means that we need to compute the non-linear matter power spectrum $P_{\rm CDM}$ in the CDM-only model using the standard \texttt{halofit} prescription.
\subsection{Datasets}
\label{ssec:dataset}
\subsubsection{Cosmic microwave background (CMB)}
We consider the CMB measurements of the most recent Planck data release~\cite{Ade:2015xua,Aghanim:2015xee},
using the full temperature power spectrum at all multipoles ($2 \leq \ell \leq 2500$, \texttt{Planck TT})
, the polarization power spectra only in the low multipoles range
($2 \leq \ell \leq 29$, \texttt{lowP}) and the \texttt{lensing} likelihood
computed from the 4-point correlation function.
Since there could be still some level of residual systematics contamination~\cite{Ade:2015xua},
we neglect the polarization measurements at high multipoles
(\texttt{highP}), following therefore a very conservative approach
which will ensure very robust limits. We refer to the \texttt{Planck TT + lowP + lensing} combination of data as the
``CMB'' dataset.

\subsubsection{KiDS data}
\label{sss:kids}
An essential point of this study is the addition of the measurements of the
Kilo Degree Survey (KiDS)~\cite{Hildebrandt:2016iqg,Joudaki:2016kym}
using the methodology explained in \cite{Hildebrandt:2016iqg}.

The KiDS data can be used to reconstruct the 2-point shear correlation functions
$\xi_\pm^{ij}(\theta)$ for the ${i,j}$  tomographic bin combination at the angle $\theta$.
The dataset that we use here is from KiDS-450~\cite{Kuijken:2015vca,Hildebrandt:2016iqg,Conti:2016gav} and
covers an effective area of $360\text{ deg}^2$. The median redshift is
$z_m=0.53$, while the effective number density is $n_{\rm eff}=8.5$ galaxies arcmin${}^{-2}$.
The experiment covers 7 angular bins in the range 0.5 to 72 arcmins for
$\xi_+^{ij}(\theta)$ and 6 angular bins between 4.2 and 300 arcmins for  $\xi_-^{ij}(\theta)$.

The calibration of the photometric redshift distributions
is made through the ``weighted direct calibration'' (DIR)
method presented in Ref.~\cite{Hildebrandt:2016iqg}.
This uses the data of external, overlapping spectroscopic
surveys and creates series of 1000 bootstrap realizations
to obtain the uncertainties and the correlations between 
the tomographic bins. Each bootstrap sample is used for a fixed number
of iterations of the MCMC scan performed here. This bootstrap procedure
ensures that the analysis is statistically unaffected by the 
photometric redshift bias corrections~\cite{Joudaki:2016kym}, 
which can instead significantly change the results of the analysis of
e.g.~CFHTLenS data and may alleviate the tension with the value
of $\sigma_8$ determined by Planck~\cite{Kitching:2016hvn}.

Furthermore, the KiDS data are analyzed taking into account the intrinsic galaxy alignments, for which
the correlations of intrinsic ellipticities of galaxies with each other and with the shear of background sources
must be considered. This is done varying two nuisance parameters: the amplitude $A_{\rm IA}$ and
the redshift dependence $\eta_{\rm IA}$ (see Ref.~\cite{Joudaki:2016mvz}).
As we have checked that our results do not change significantly if we
turn on these ``extended systematics'' settings~\cite{Joudaki:2016kym}, 
we will only show the results obtained within the standard prescription, i.e.~$\eta_{\rm IA}=0$ and $-6\le A_{\rm IA}\le6$.

To perform the analyses as presented by the KiDS collaboration~\cite{Hildebrandt:2016iqg,Joudaki:2016kym},
we should include the calculation of baryonic effects in the
nonlinear matter power spectrum, that are computed using  \texttt{HMCODE}~\cite{Mead:2015yca}.
As we discussed in Subsec.~\ref{ssec:dm}, this code gives biased results
when applied to the (unusual) MDM cosmology when the non-cold dark matter temperature is much smaller than the standard neutrino one.
Since we are using \eq{eq:pmdm} for computing the non-linear
matter power spectrum, we shall not use \texttt{HMCODE} and the related
prescriptions on the baryonic feedback.

\subsubsection{Satellite galaxies}
\label{sss:sat}
As we have already introduced,
one of the problems of the \lcdm~model at galactic and sub-galactic scales is the one of the missing satellite galaxies.
Here we explain how we compute the constraints from the observations
of dwarf satellite galaxies in the Milky Way (MW).

Dwarf galaxies are usually faint and small objects
that must be observed and correctly identified as satellites of the
MW. In the following we briefly comment on the current observational status~\cite{Schneider:2016uqi, Polisensky:2010rw}.
The number of known standard satellites of our galaxy is eleven.
The SDSS experiment, with a sky coverage of $f_{\rm sky}\simeq 0.28$,
observed other fifteen satellites~\cite{Wolf:2009a},
with a corresponding poissonian error of $\sim4$. 
Recently, the Dark Energy Survey (DES) has reported the discovery of
eight new Milky Way companions, which could potentially be ultra-faint
satellite galaxies~\cite{Bechtol:2015cbp,Drlica-Wagner:2015ufc}. In the absence
of a robust confirmation of the fact that these new eight DES candidates are truly
all of them Milky-Way satellite galaxies, we adopt here the conservative approach of Ref.~\cite{Schneider:2016uqi}
and restrict ourselves to the classical satellites plus the
extrapolated number of the SDSS measurements, see the discussion that follows.

We assume here that the number within the SDSS footprint is binomially distributed
with probability $f_{\rm sky}\simeq 0.28$, following an isotropic
distribution. Then, a SDSS-like experiment would have observed
$54$ MW satellite galaxies over the entire sky, with an
associated (binomial distribution) error of $11$. Therefore, together with the eleven classical MW
satellites, this would imply a total number of $N_{\rm sat}=65\pm 11$. Following
Ref.~\cite{Schneider:2016uqi}, we assume a halo-to-halo
scatter~\cite{BoylanKolchin:2009an} by reducing our
estimated number of MW satellite galaxies, quoted above, by $10\%$.
Nevertheless, we are aware that this number is probably an underestimate
of the true number of dwarf satellites, as a consequence of the
technical challenges of the observation. 
Notice however that, due to the incompleteness of the SDSS sample, it could be possible that,
when accounting for corrections in e.g.\ the observed luminosity function,
the number of satellite galaxies around the MW could be much larger
than these estimates, see e.g.\ Ref.~\cite{Tollerud:2008ze}.
For this reason, we present the results obtained combining the satellites likelihood with other
datasets following two different approaches.
In the most conservative
scenario, labeled ``SAT(low)'' in the following, we apply the observed number of MW dwarf
satellite galaxies $\nsat^{\rm obs}= 58\pm 11$ as a lower limit only, by
means of a half-gaussian likelihood, following
e.g.~Ref.~\cite{Diamanti:2017xfo}.
In this conservative approach we
apply the dwarf galaxy bounds only when the
number of satellite galaxies predicted within a given model is below the mean number of satellite galaxies
that are observed. In this way we envisage the putative situation in
which not all dwarf spheroidal galaxies have been detected, being the
current estimates subject to increase by ongoing and/or future searches. 
We also follow a more aggressive scenario in which we apply the current
measurement $\nsat^{\rm obs}= 58\pm 11$ via a standard gaussian
likelihood. This latter case will be referred to as ``SAT''.

Another problem related to dwarf satellites is that it is also difficult to infer the mass of these objects,
since they are dominated by dark matter and the only possibility to measure
the properties of the DM halo is through stellar kinematics
inside the object.
Studies that use different profiles for the halo suggest that
all the known dwarfs have a mass larger than $M_{\rm min}=10^8\, h^{-1} M_\odot$~\cite{Brooks:2012a},
a number that we shall use in the calculations explained below.

For the theoretical computation of the number of satellites
we follow the procedure described
in Refs.~\cite{Schneider:2014rda,Schneider:2016uqi,Diamanti:2017xfo},
based on a conditional mass function that is
normalized taking into account the results of the N-body simulations.
The function which gives the expected number of
dwarf satellite galaxies with a given mass $M_{s}$ reads as~\cite{Schneider:2014rda, Schneider:2016uqi}:
\be
  \label{eq:sat}
  \frac{d\nsat}{d \ln M_{s}}
  =
  \frac{1}{C_n}
  \, \frac{1}{6 \pi^2}
  \,
  \left(
    \frac{M_{h}}{M_{s}}
  \right)
  \, \frac{ P(1/R_{s}) }{R^3_{s}
  \sqrt{2\pi(S_{s} - S_{h})}}
  \,~.
\ee
Here $P(1/R_s=k)$ is the linear matter power spectrum
for the given cosmological model, $h$ stands for ``host halo'',
i.e.~the MW galaxy, and the subscript $s$ stands for ``satellite'' or ``subhalo''.
The coefficient $C_n = 45$ is chosen to mimic the results of N-body simulations~\cite{Lovell:2013ola}.
In our calculations we assume $M_{h}=1.77 \cdot 10^{12} h^{-1}M_\odot$ for the MW.
This may be not the exact mass of our galaxy,
but it is chosen because it lies in the estimated range for the MW mass~\cite{Guo:2009a}
and it matches the mass of the
Aquarius-D2 simulation~\cite{Springel:2008cc},
from which the calibration on the N-body simulations results is computed~\cite{Lovell:2013ola}.
The estimated number of satellites $\nsat^{\rm th}$
is obtained integrating
the above~\eq{eq:sat} between the minimum mass of
satellites $M_{\rm min}$, previously described,
and the mass of the host halo $M_{h}$:
\be\label{eq:nsat}
  \nsat^{\rm th}
  =
  \int_{M_{\rm min}}^{M_{h}}
  \frac{d\nsat}{d \ln M_{s}}\, dM_{s}
  \,.
\ee

The parameters describing the subhalo (or the host halo)
are the radius $R_s$ ($R_h$),
the mass $M_s$ ($M_h$) and
the corresponding variance $S_s$ ($S_h$).
They are related by:
\bea\label{eq:s_halo}
  S_i(M)
  &=&
  \frac{1}{2\pi^2}
  \int_0^\infty k^2 P(k) W^2(k|M) dk
  \,,\\ \label{eq:m_halo}
  M_i
  &=&
  \frac{4\pi}{3} \, \Omega_m \rho_c (c R_i)^3
  \,~,
\eea
where $i=s, h$.
The parameter $c = 2.5$ is used to calibrate the calculation on 
N-body simulation results.

To evaluate the variance, we use a procedure based on a re-derivation
of the Press \& Schechter~\cite{Press:1973iz} mass function,
for which we use a \emph{k-sharp filter} approach~\cite{Lovell:2015psz}
to cut all the scales $k$ below the cut-off scale $k_s=1/R_{s}$, and
$R_s$ depends on the subhalo mass as in \eq{eq:m_halo}.
This filter is written in terms of a
window function
\begin{equation}
  W(k|M) =
  \begin{cases}
    1, \quad \text{if} \quad k \leq k_s(M)  \,;\\
    0, \quad \text{if} \quad k > k_s(M) \,.\\
  \end{cases}
\end{equation}

\subsection{Numerical analyses}
\label{ssec:theo}
We base our analyses on the standard \lcdm~model,
with the addition of a second non-cold dark matter component.
The parameters that we vary in our analyses are then
the energy density of baryons, $\omega_b\equiv\Omega_b h^2$;
the total energy density of the dark matter components $\omega_{DM}$,
the fraction of the total dark matter mass-energy density in the form of non-cold dark
matter ($f_X$) and its temperature
through the logarithm $\log_{10}(T_X/T_\nu)$
(see also Subsec.~\ref{ssec:dm});
the optical depth to reionization, $\tau$; the ratio of the sound
horizon to the angular diameter distance at decoupling $\Theta$; and
the amplitude and the tilt of the primordial power spectrum of curvature
perturbations, $\ln(10^{10}A_s)$ and $n_s$.

Other cosmological parameters are fixed to their standard values
as follows. The sum of the three active neutrino masses is fixed to zero.
Despite neutrino oscillation measurements tell us
that at least two of the neutrinos must have a mass, with a minimum
total mass of
$\sum m_\nu\simeq0.06$~eV for the normal ordering, the error that we make when fixing $\sum m_\nu\simeq0$ is small.
For the three massless neutrinos, we fix the standard value
$\neff^{\nu} = 3.046$, corresponding to the three active neutrino contribution 
obtained in the limit of non-instantaneous neutrino
decoupling~\cite{Mangano:2005cc,deSalas:2016ztq}.

For the Planck CMB and satellite galaxy number counts we use the
Markov Chain Monte Carlo (MCMC) tool Monte Python~\cite{Audren:2012wb}, interfaced
with the Boltzmann solver CLASS~\cite{Lesgourgues:2011re}.
We use then the obtained covariance
matrices to run MCMC with KiDS data, by means of the Boltzmann solver
\camb~\cite{Lewis:1999bs} together with its MCMC companion
\cosmomc~\cite{Lewis:2002ah}~\footnote{For the latter one, we use the July 2015 version with the required
modifications  to perform the analyses of the KiDS
data~\cite{Joudaki:2016kym}, publicly available at
\url{http://github.com/sjoudaki/kids450}.}.

Together with the cosmological parameters, we vary all the required nuisance parameters involved in the Planck and the KiDS likelihoods.

\section{Results}
\label{sec:results}
\begin{figure*}[t]
  \begin{center}
    \hspace{-1cm}
    \begin{tabular}{cc}
      \includegraphics[width=6cm]{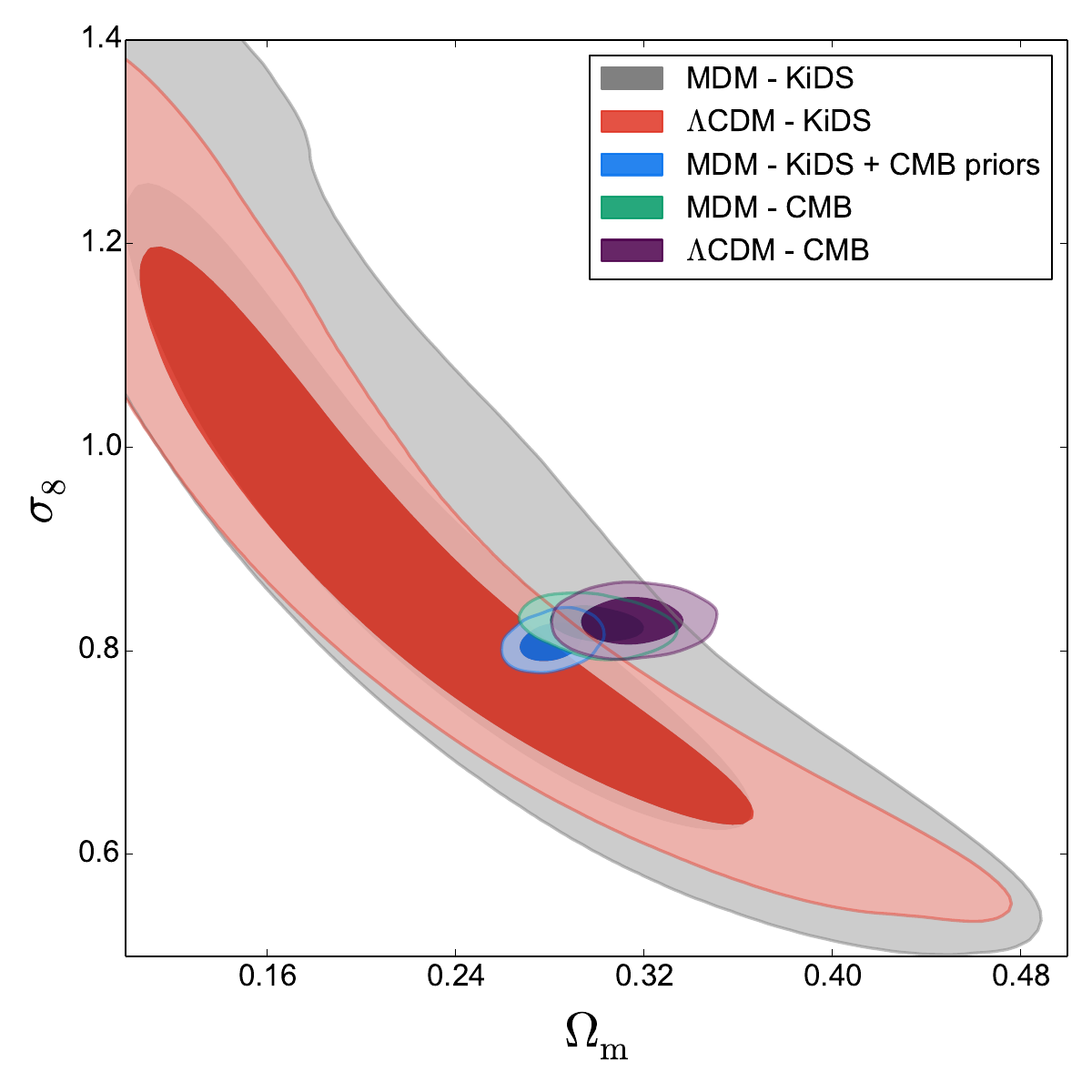}&\includegraphics[width=6cm]{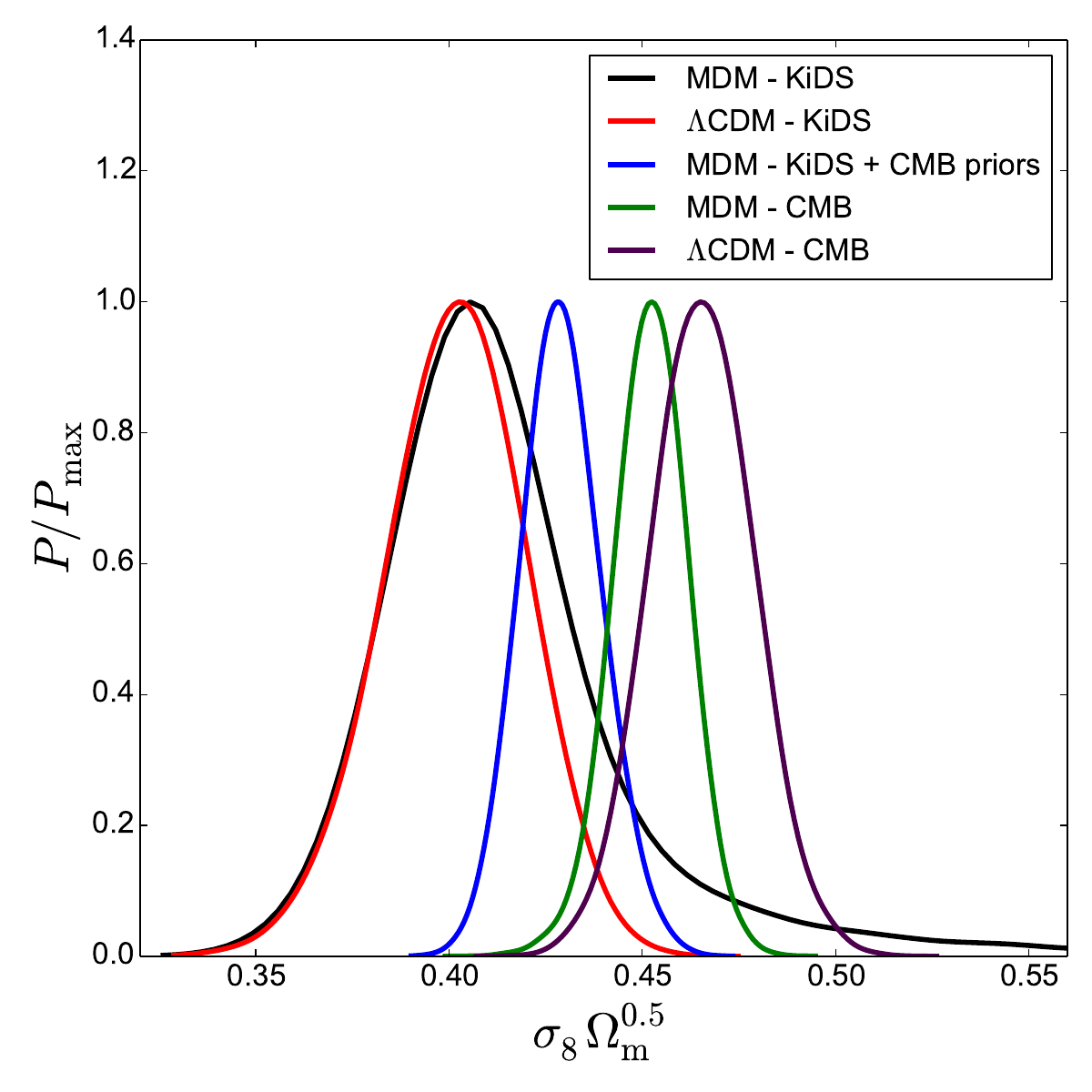}
    \end{tabular}
  \end{center}
  \caption{Left panel: 68\% and 95\% CL allowed contours in the
    ($\Omega_m$, $\sigma_8$) plane, see text for details. Right panel: one-dimensional posterior probability
distribution of the $S_8$ quantity for the same data combinations and models shown in the left panel.}
    \label{fig:omegams8}
\end{figure*}

Figure~\ref{fig:omegams8}, left panel, depicts the 68\% and 95\% CL allowed contours in the
($\Omega_m$, $\sigma_8$) plane resulting from a number of possible
data combinations and comparing different underlying cosmological
models. In the simplest $\Lambda$CDM picture, the allowed regions for the
KiDS and Planck datasets do not basically overlap, showing a clear
tension, as pointed out before by Refs.~\cite{Hildebrandt:2016iqg,Joudaki:2016kym}. Such a tension is
clearly alleviated when one considers the possible MDM extension to
the minimal $\Lambda$CDM picture: notice that the contours for KiDS
and Planck overlap for a larger region in the MDM case. This very same improvement
can be noticed from the one-dimensional posterior probability distribution of the
$S_8=\sigma_8\sqrt{\Omega_m}$ quantity in the right panel of
Fig.~\ref{fig:omegams8}. 

A possible way of quantifying the tension in the
measurements of $S_8$ arising from the Planck and KiDS datasets in possible extensions of the
$\Lambda$CDM framework has been presented and used in
Ref.~\cite{Joudaki:2016kym}.
The tension is defined by:
\begin{equation}
T(S_8)
\equiv
|\bar{S}^{D_1}_8-\bar{S}^{D_2}_8|
/\sqrt{\sigma^2(S^{D_1}_8)+\sigma^2(S^{D_2}_8)}~,
\end{equation}
where $D_1$ and $D_2$ refer to the Planck and KiDS datasets,
$\bar{S}_8$ is the mean value over the posterior distribution and
$\sigma$ refers to the $68\%$~CL error on $S_8$.
If we compute the value of $T(S_8)$ from the constraints obtained within the MDM scenario explored here, we get a
displacement of $T^{\textrm{MDM}}(S_8) \simeq 1\sigma$,
which indicates that the $2.3\sigma-2.8\sigma$~%
\footnote{The authors of \cite{Hildebrandt:2016iqg} quote a
  $2.3\sigma$ tension among the obtained $S_8$ values
  from Planck (CMB temperature and low-$\ell$ polarization) and KiDS.
  Our analysis of the KiDS results with the prescriptions
  published together with Ref.~\cite{Joudaki:2016kym} leads to a value of
  $T(S_8)\simeq2.8$, with small variations when we consider
  our results from Planck CMB data or 
  the Planck chains publicly available at \url{https://wiki.cosmos.esa.int/planckpla}.
  When the full CMB dataset we explore here is considered, which includes the Planck lensing likelihood,
  the tension shifts to $T(S_8)\simeq2.5$.}
  obtained between Planck and KiDS $S_8$ values within the
canonical $\Lambda$CDM scenario is considerably reduced.
The level of the improvement is very similar
to that reached when other possible extensions to the minimal
scenario are considered, as, for instance, in the presence of a dark
energy equation of state or within modified gravity
models~\cite{Joudaki:2016kym}.
The fact that assuming a MDM scenario the tension
between the Planck and KiDS constraints is
strongly alleviated fully justifies the combination of these
two datasets, already depicted in both panels of Fig.~\ref{fig:omegams8}.

\begin{figure*}[t]
  \begin{center}
    \hspace{-1cm}
    \begin{tabular}{ccc}
      \includegraphics[width=6cm]{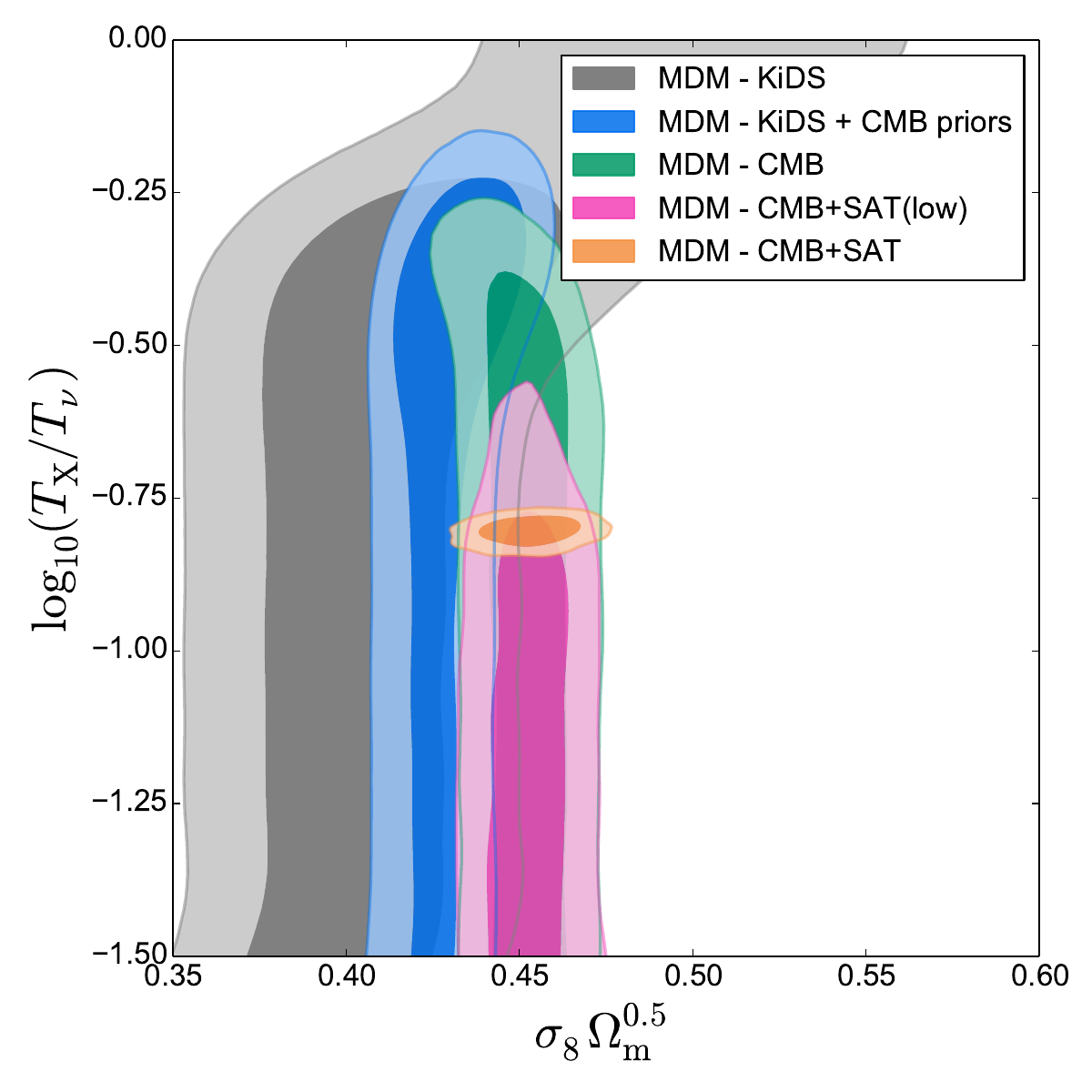}&\includegraphics[width=6cm]{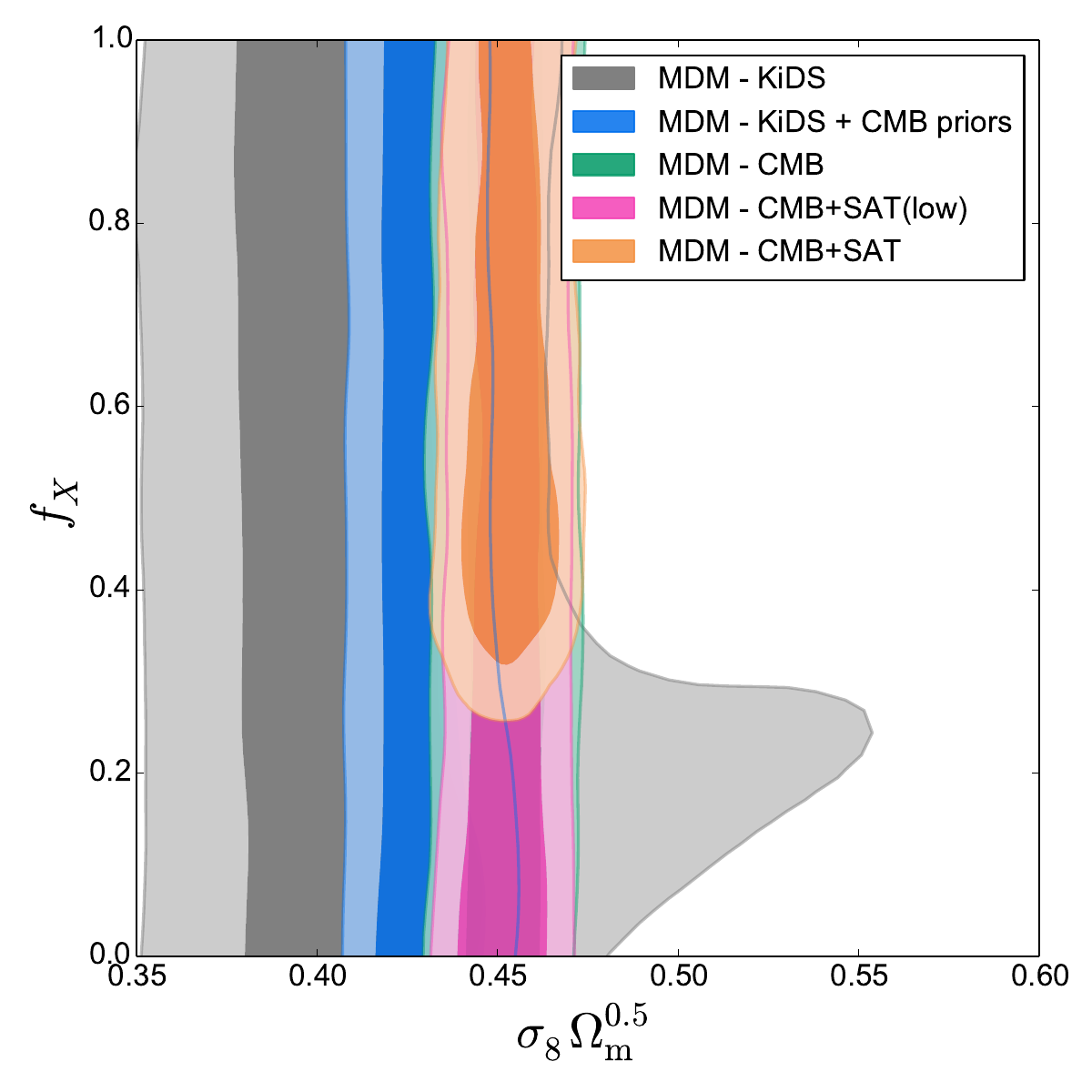}
    \end{tabular}
  \end{center}
  \caption{68\% and 95\% CL allowed contours in the ($S_8$,
$T_X/T_\nu$) and ($S_8$, $f_X$) planes resulting from different datasets:
KiDS alone, CMB alone, the combination of KiDS and
CMB data, 
and the combination of CMB measurements with
MW satellite number counts, see text
for details.}
    \label{fig:s8sat}
\end{figure*}

Figure~\ref{fig:s8sat} shows the two-dimensional 68\% and 95\% CL 
allowed contours in
the ($S_8$, $T_X/T_\nu$)
and ($S_8$, $f_X$) planes. We illustrate the results from KiDS, CMB and their
combination.
We can notice from the contours shown in the left panel of Fig.~\ref{fig:s8sat}
that there exists a degeneracy between the $S_8$ quantity and
$T_X/T_\nu$.
The reason for that is because the
larger $T_X$ is, the closer the non-cold dark matter component behaves
as a hot dark matter fluid, and therefore a larger matter component
(and, consequently, a larger value of $S_8$) would be required to compensate for the suppression of perturbations
at small scales. As expected, there are no bounds on the non-cold dark
matter fraction $f_X$, as one can see in the right panel of
Fig.~\ref{fig:s8sat}. Indeed, for a very small value of $T_X$ the non-cold
dark matter is observationally indistinguishable from a pure cold
component and therefore its relative abundance is perfectly
compatible with $f_X=1$. 

As previously stated, the motivation for MDM scenarios is twofold. We
have already shown above that a non-cold dark matter component 
provides a possible way of alleviating the tension between Planck CMB
measurements and tomographic weak lensing data from KiDS. 
As mentioned in our introductory section, there is another very important motivation to look for
additional extensions of the minimal cold dark matter picture, namely,
the so-called small scale crisis of the $\Lambda$CDM, and, in
particular, the \emph{Milky Way satellite
problem}~\cite{Klypin:1999uc, Moore:1999nt}. We have
therefore also considered in our analyses the constraints from the MW
dwarf satellite galaxies, that we are going to discuss in what follows.

\begin{figure*}[t]
  \begin{center}
         \includegraphics[width=0.98\textwidth]{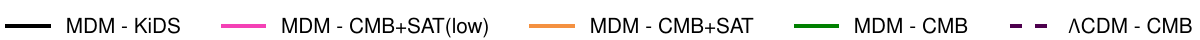}
     \includegraphics[width=0.98\textwidth]{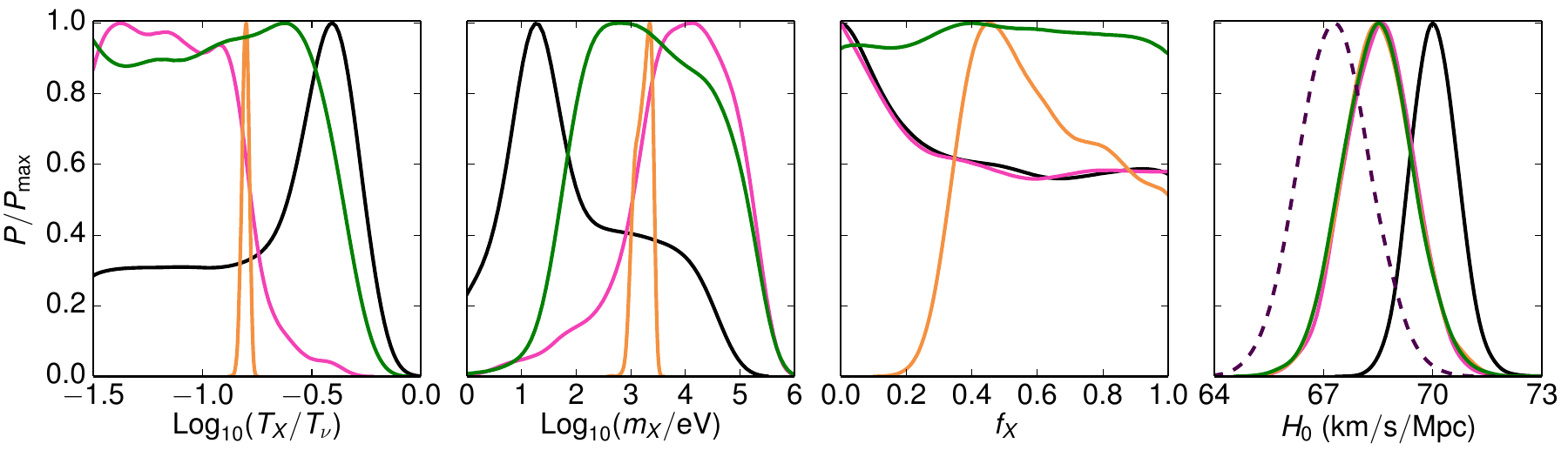}
  \end{center}
  \caption{One-dimensional posterior probability distribution for the
    $T_X/T_\nu$ ratio, for the non-cold dark matter mass $m_X$, its relative fraction $f_X$ and for the
    Hubble constant $H_0$ for 
different data combinations in the MDM scenario, see the text for
details. For comparison purposes, in the case of
the Hubble constant, we also depict the resulting one-dimensional
distribution from a fit to Planck CMB data in the context of a $\Lambda$CDM cosmology.}
    \label{fig:1dfigs}
\end{figure*}

The first panel of Fig.~\ref{fig:1dfigs} shows the one-dimensional 
probability  distribution of the non-cold dark matter temperature
relative to that of the neutrino bath $T_X/T_\nu$. Planck CMB data measurements provide the $95\%$~CL upper bound of
$T_X/T_\nu<0.4$, which could be naively translated into a contribution to
$\Delta\neff=(T_X/T_\nu)^4\simeq 0.02$ during the periods
in the universe expansion history in which these non-cold dark matter
particles are relativistic. Notice that this value is much smaller than the limit obtained from Planck CMB measurements on the
contribution from (massless) dark radiation or light sterile neutrinos.
However, the limits are not directly comparable as the energy density of the non-cold dark matter
component explored here is that of a non-relativistic fluid for a large region of the
parameter space. KiDS measurements show a preference for higher temperatures,
indeed the $95\%$~CL upper limit
is set by the upper prior in the $T_X/T_\nu$ ($T_X/T_\nu<1$) parameter.
When analyzing KiDS data taking into account
the Planck CMB constraints on all the cosmological parameters, we obtain $T_X/T_\nu<0.6$ at $95\%$~CL.
We also show in Fig.~\ref{fig:1dfigs} the results obtained
combining the MW satellites counts with CMB measurements.
The blue curve depicts the one-dimensional 
probability  distribution of the non-cold dark matter temperature when
the current estimates of MW satellite galaxies are treated as regular gaussian
priors and are combined with Planck CMB measurements. This data
combination constrains the temperature to lie within a narrow region, 
favoring scenarios with values of $T_X/T_\nu$ in the range $0.15<
T_X/T_\nu<0.17$, therefore smaller than those quoted previously for the KiDS
plus Planck MDM case.
This difference can be explained
in terms of the preferred value of $m_X$, which turns out to be larger
for Planck plus MW satellites than for Planck plus KiDS data. If we
instead consider the MW likelihood in the form of a conservative
half-gaussian likelihood, imposing the current observed number of galaxies only as
a lower limit, there is no lower bound on the non-cold dark matter
temperature, as the half-gaussian likelihood turns out to be perfectly 
compatible with a pure $\Lambda$CDM universe, for which the number of satellite galaxies is around 160.
The upper limit on
$T_X/T_\nu$ is very close to the one quoted above for our more
aggressive MW likelihood scenario ($T_X/T_\nu<0.19$ at $95\%$~CL), and 
smaller also than the constraint obtained from Planck plus KiDS data.

The second panel of Fig.~\ref{fig:1dfigs} shows the one-dimensional
probability distribution of the non-cold dark matter mass
$m_X$~\footnote{We recall that $m_X$ is a derived parameter in our
  analyses.}. Notice that there exist a $95\%$~CL lower limit from
Planck data on the mass of the MDM component $m_X>32$~eV. This bound on the mass is related to the fact
that, below that region, the non-cold dark matter fluid behaves as a \emph{hot} dark matter
component (even if its temperature is lower than the neutrino one) and CMB observations do not allow for such large
contribution from hot dark matter relics. On the other hand, KiDS
measurements show a mild preference for a dark
matter mass in the sub-keV region, that can provide the suppression
of the small-scale perturbations required
to reduce the value of the $S_8$ quantity for KiDS.
The combination of Planck CMB plus KiDS
weak lensing data does not significantly change these findings.

When considering the MW dwarf
satellite constraints in their less conservative form, we observe that there is
a preferred narrow region for the non-cold dark matter mass, which, as we shall further
illustrate in short, turns out to be very close to the warm dark
matter region~\footnote{For pioneer work on WDM cosmologies see
  Refs.~\cite{Moore:1999gc,Bode:2000gq,AvilaReese:2000hg,Narayanan:2000tp,Viel:2005qj}.}. The mass of the particle lies in the range
$0.95$~keV$<m_X<2.9 $~keV ($95\%$~CL). Additional and independent constraints from power spectrum 
measurements from the Lyman alpha forest flux and from the universe
reionization history can be applied in this case, see
Refs.~\cite{Viel:2013apy,Baur:2015jsy,Irsic:2017ixq,Yeche:2017upn,Lopez-Honorez:2017csg}
for the most recent analyses. 

When adopting the more conservative, half-gaussian approach for
the MW likelihood, the lower bound we get for the non-cold dark
component is very close to the region quoted above, $m_X>0.09$~keV at
$95\%$~CL, as lower values of the non-cold dark matter mass will lead to a very large
suppression of the matter power spectrum at galactic and sub-galactic
scales, in which case the number of MW satellite galaxies gets
strongly reduced. However, for this case, there is
no upper limit on $m_X$, as a model with only the cold dark matter component is perfectly allowed
once the upper bound in the observed number of MW galaxies is no
longer considered, largely relaxing the allowed mass range. 

Notice that the preferred region for the mass $m_X$ obtained considering the MW
satellites observations is larger than the one suggested by KiDS weak
lensing measurements. This is due to the fact that the suppression in
the growth of structure required to satisfy MW satellite galaxy
observations is associated to a smaller scale (large wavenumber $k$)
than the one required to explain KiDS weak lensing data. 
The differences in the preferred values of $m_X$ from weak lensing and
from KiDS lead to differences in the allowed values of $T_X/T_\nu$. As
previously stated, the bound on $m_X$ directly depends on the constraint in
the abundance of the particle. Consequently, the larger (smaller)
the allowed mass is, the smaller (larger) the temperature should be to
satisfy CMB constraints, see Eq.~(\ref{eq:mwdm}). 

The bounds on the non-cold dark matter fraction are shown in the third
panel of Fig.~\ref{fig:1dfigs}. Planck measurements result in an almost
completely flat distribution for $f_X$, as particles with small
temperatures and large masses will produce CMB photon temperature and
lensing patterns that are identical to those obtained in the pure cold dark matter
case. KiDS and its combination with CMB
measurements lead also to flat distributions for $f_X$, with limits coinciding with
the assumed priors on $f_X$.
When dealing with the MW satellites
likelihood in the less conservative approach, its combination with Planck CMB sets a robust preference
for $f_X>0$, i.e.~$f_X>0.34$ at $95\%$~CL, while in the more
conservative approach the $f_X$ distribution is flat.  
Therefore, the results that we obtain
when considering the MW dwarf satellites in the less conservative approach followed here
strongly suggest the need for a non-zero non-cold dark matter
component. The explanation is simple:
while we observe a number around 60 satellite galaxies,
the corresponding number for a CDM-only case would be $\sim160$.
A suppression of the matter perturbations at small scales
(as, for instance, that associated to a non-cold dark matter
component) is required, in order to match the predicted number and the observed one.

We can also note from the two panels of Fig.~\ref{fig:s8sat} that the allowed contours in the ($S_8$,
$T_X/T_\nu$) and ($S_8$, $f_X$) planes after considering the MW
dwarf satellite galaxies likelihood do not deviate significantly from the CMB constraints.
This shows that the second dark matter component required to fully explain the satellites counts
at galactic scale does not help in solving the tension in the $S_8$ parameter
that exists between Planck CMB and KiDS weak lensing data.
On the other hand, there exists also the possibility
that the current measures of the number of MW dwarf galaxies
are only underestimations of the true number,
and future observations will increase the present statistics.
Figure~\ref{fig:s8sat} illustrates also such a
possibility, when combining MW satellite number counts with CMB data.
In this more conservative approach to the MW satellites problem,
we can notice that the satellites limits on $f_X$ overlap with those from the CMB,
adding no extra information on the non-cold dark
matter abundance.

Finally, the fourth panel of Fig.~\ref{fig:1dfigs} shows the
one-dimensional posterior distribution of the Hubble constant $H_0$.
Notice that, as in the case of Ref.~\cite{Joudaki:2016kym} for
other possible extensions of the $\Lambda$CDM scheme, the
Hubble parameter values show a better agreement with direct
estimates of $H_0$~\cite{Efstathiou:2013via,Riess:2016jrr} in the MDM
scheme than in the pure cold dark matter scenario. The Planck constraint on
the Hubble constant in the MDM scenario explored here
($H_0=68.5\pm 0.9$~km~s$^{-1}$~Mpc$^{-1}$) is higher than in the
pure $\Lambda$CDM case ($H_0=67.3\pm 1$~km~s$^{-1}$~Mpc$^{-1}$),
i.e.~the mean value of $H_0$ is shifted by approximately $1.5\sigma$. 
The reason for the larger preferred value of $H_0$ in the context of MDM
scenarios can be understood as follows. From Fig.~\ref{fig:omegams8}
it is straightforward to infer that the value of $\Omega_m$ in these scenarios is generically lower
than in the pure $\Lambda$CDM case. This fact is reinforced when
combining Planck and KiDS data. As the CMB peaks structure does
not leave much freedom on the value of the physical (total) matter
energy density $\Omega_m h^2$, a smaller value of
$\Omega_m$ requires the mean value of $H_0$ to be larger, in order to
leave the product $\Omega_m h^2$ unchanged.
When combining Planck CMB and KiDS
datasets we obtain $H_0=70.0 \pm 0.6$~km~s$^{-1}$~Mpc$^{-1}$, 
showing a much better matching to the Hubble parameter value extracted from local
observations. This value, for instance, is consistent (within
$1\sigma$) with that of Ref.~\cite{Efstathiou:2013via}, $H_0=72.5
\pm 2.5$~km~s$^{-1}$~Mpc$^{-1}$, and the $3.4\sigma$ tension within
the $\Lambda$CDM paradigm between CMB and the Hubble constant estimates
of Ref.~\cite{Riess:2016jrr}, $H_0=73.24 \pm 1.74$~km~s$^{-1}$~Mpc$^{-1}$, is reduced to the $2\sigma$ level. 
Finally, the combination of MW satellites and
Planck data leaves unchanged the value of $H_0$ from CMB alone in MDM
scenarios, as expected, since MW satellites observations do not
require a change on the dark matter abundance but on its nature, and
therefore no shift is required in $H_0$.

\section{Conclusions}
\label{sec:conclusions}
Observations at galactic and sub-galactic scales compromise the
viability of the canonical $\Lambda$CDM paradigm, which otherwise 
provides an excellent fit to large scale structure and Cosmic
Microwave Background (CMB) observations. The number of satellite galaxies in
Milky-Way (MW) sized halos and the measured kinematics of the MW satellites
pose the question of whether a universe made out of
a pure cold dark matter component and a dark energy fluid can
successfully explain all cosmological
observations. Furthermore, there are a number of additional
inconsistencies among small scale predictions from the \lcdm\ model and observations from recent data
releases from tomographic weak gravitational lensing surveys, as those from the Kilo Degree-450 deg$^2$
Survey (KiDS-450)~\cite{Hildebrandt:2016iqg,Joudaki:2016kym}. More
concretely, there are discrepancies in the value of
the amplitude of the density fluctuations at a given matter density $\Omega_m$, commonly quantified in terms of
$S_8=\sigma_8\sqrt{\Omega_m}$. 

Models with an admixture of cold and non-cold dark
matter particles (MDM models) may potentially alleviate the
$\Lambda$CDM observational problems outlined above, as the free
streaming nature associated to a non-totally cold dark matter
component will suppress the matter power spectrum on the smallest
scales, leading to a better agreement among large and galactic
and sub-galactic scales measurements. Here we have analyzed these MDM
scenarios using the most recent tomographic weak lensing measurements from the KiDS-450
survey, combining them with Planck CMB data. We have also studied the constraints derived using the
current estimates for the observed number of MW satellite galaxies. 

In a similar way to other extended cosmological
models~\cite{Joudaki:2016kym}, the tension in the measurements from
Planck and KiDS of the $S_8$ quantity is reduced from $2.3-2.8\sigma$ to
$1\sigma$ in MDM scenarios. Furthermore, the value of the Hubble
parameter $H_0$ is perfectly consistent with that measured by late
universe observations~\cite{Efstathiou:2013via}. We find $H_0=70.0 \pm 0.6$~km~s$^{-1}$~Mpc$^{-1}$ after combining Planck CMB and KiDS
tomographic weak lensing data, value which is in a good agreement with the Hubble parameter value extracted from local
observations, see e.g. Ref.~\cite{Efstathiou:2013via}, which quotes a
value of $H_0=72.5\pm 2.5$~km~s$^{-1}$~Mpc$^{-1}$. 

We have also searched for the allowed ranges on the non-cold dark matter
properties, as its temperature and its mass. We find a $95\%$~CL
upper limit $T_X/T_\nu<0.6$ after combining Planck CMB and KiDS
data. Current estimates of the number of satellite galaxies,
$\nsat^{\rm obs}= 58\pm 11$, when translated into a standard gaussian
likelihood and combined with CMB measurements, prefer smaller values
of the temperatures ratio ($0.15<T_X/T_\nu<0.17$ at $95\%$~CL), as they would require larger values of $m_X$
to suppress the growth of structure at the scales involved in MW halo 
observations. However, if current MW satellite observations are
conservatively interpreted as a half-gaussian likelihood, imposing
the current measured number only as a lower limit, the lower bound on
the non-cold dark matter temperature disappears and we obtain $T_X/T_\nu<0.19$ at $95\%$~CL. Concerning the non-cold dark matter mass and its abundance,
after combining Planck CMB and KiDS measurements, we find a (mild)
preference for a sub-keV non-cold dark matter particle mass, with no particular
evidence for a non-zero abundance of such a component. However, satellite
counts data, in their more aggressive interpretation and combined with
CMB measurements, isolate the preferred regions $0.95$~keV$<m_X<2.9 $~keV and $f_X>0.34$ at $95\%$~CL, robustly establishing the need
for the existence of such a keV warm dark matter particle. In the
most conservative approach there is not such a preference for
$f_X\neq0$ and only a lower bound on the non-cold dark matter
mass exists, $m_X>0.08$~keV at $95\%$~CL. 

Therefore, a sub-dominant, non-cold dark matter
component with $m_X\sim$~keV could in principle alleviate some existing
tensions between CMB and low redshift observations. However, the
masses and temperatures required to explain weak lensing and MW observations are
rather different. While the scale of the power spectrum suppression
required by KiDS data needs a sub-keV non-cold dark
matter mass and a temperature $T_X\sim T_\nu/2$, the ones required by
Milky Way satellites observations are associated to larger masses and smaller temperatures. 
Future weak lensing and MW satellites observations will further sharpen the
preferred regions, either enlarging or diminishing
the existing differences among the current weak lensing and MW
preferred constraints for the non-cold dark matter temperature and
mass. Future work with simulated data will be devoted elsewhere
to further corner MDM scenarios.

\begin{acknowledgments}
The authors would like to thank Francisco Villaescusa-Navarro and Alexander Mead for useful discussions
on the nonlinear power spectrum parameterization
and Shahab Joudaki for details on the KiDS analysis.
The work of S.G.~was supported by the Spanish grants
FPA2014-58183-P,
Multidark CSD2009-00064 and
SEV-2014-0398 (MINECO),
and PROMETEOII/2014/084 (Generalitat Valenciana).
O.M.~is supported by PROMETEO II/2014/050,
by the Spanish Grant FPA2014--57816-P of the MINECO,
by the MINECO Intramural OEP2010,
by the MINECO Grant SEV-2014-0398 and
by the European Union's Horizon 2020 research and innovation programme
under the Marie Sk{\l o}dowska-Curie grant agreements 690575 and
674896. The work of R.D. was supported by NWO through two Vidi grants and partly by
University of Amsterdam. M.E. is supported by Spanish Grant FPU13/03111 of 6MECD.
\end{acknowledgments}

%


\begin{thebibliography}{78}%
\makeatletter
\providecommand \@ifxundefined [1]{%
 \@ifx{#1\undefined}
}%
\providecommand \@ifnum [1]{%
 \ifnum #1\expandafter \@firstoftwo
 \else \expandafter \@secondoftwo
 \fi
}%
\providecommand \@ifx [1]{%
 \ifx #1\expandafter \@firstoftwo
 \else \expandafter \@secondoftwo
 \fi
}%
\providecommand \natexlab [1]{#1}%
\providecommand \enquote  [1]{``#1''}%
\providecommand \bibnamefont  [1]{#1}%
\providecommand \bibfnamefont [1]{#1}%
\providecommand \citenamefont [1]{#1}%
\providecommand \href@noop [0]{\@secondoftwo}%
\providecommand \href [0]{\begingroup \@sanitize@url \@href}%
\providecommand \@href[1]{\@@startlink{#1}\@@href}%
\providecommand \@@href[1]{\endgroup#1\@@endlink}%
\providecommand \@sanitize@url [0]{\catcode `\\12\catcode `\$12\catcode
  `\&12\catcode `\#12\catcode `\^12\catcode `\_12\catcode `\%12\relax}%
\providecommand \@@startlink[1]{}%
\providecommand \@@endlink[0]{}%
\providecommand \url  [0]{\begingroup\@sanitize@url \@url }%
\providecommand \@url [1]{\endgroup\@href {#1}{\urlprefix }}%
\providecommand \urlprefix  [0]{URL }%
\providecommand \Eprint [0]{\href }%
\providecommand \doibase [0]{http://dx.doi.org/}%
\providecommand \selectlanguage [0]{\@gobble}%
\providecommand \bibinfo  [0]{\@secondoftwo}%
\providecommand \bibfield  [0]{\@secondoftwo}%
\providecommand \translation [1]{[#1]}%
\providecommand \BibitemOpen [0]{}%
\providecommand \bibitemStop [0]{}%
\providecommand \bibitemNoStop [0]{.\EOS\space}%
\providecommand \EOS [0]{\spacefactor3000\relax}%
\providecommand \BibitemShut  [1]{\csname bibitem#1\endcsname}%
\let\auto@bib@innerbib\@empty
\bibitem [{\citenamefont {Ade}\ \emph {et~al.}(2016)\citenamefont {Ade} \emph
  {et~al.}}]{Ade:2015xua}%
  \BibitemOpen
  \bibfield  {author} {\bibinfo {author} {\bibfnamefont {P.~A.~R.}\
  \bibnamefont {Ade}} \emph {et~al.} (\bibinfo {collaboration} {Planck}),\
  }\href {\doibase 10.1051/0004-6361/201525830} {\bibfield  {journal} {\bibinfo
   {journal} {Astron. Astrophys.}\ }\textbf {\bibinfo {volume} {594}},\
  \bibinfo {pages} {A13} (\bibinfo {year} {2016})},\ \Eprint
  {http://arxiv.org/abs/1502.01589} {arXiv:1502.01589 [astro-ph.CO]}
  \BibitemShut {NoStop}%
\bibitem [{\citenamefont {Bertone}\ \emph {et~al.}(2005)\citenamefont
  {Bertone}, \citenamefont {Hooper},\ and\ \citenamefont
  {Silk}}]{Bertone:2004pz}%
  \BibitemOpen
  \bibfield  {author} {\bibinfo {author} {\bibfnamefont {G.}~\bibnamefont
  {Bertone}}, \bibinfo {author} {\bibfnamefont {D.}~\bibnamefont {Hooper}}, \
  and\ \bibinfo {author} {\bibfnamefont {J.}~\bibnamefont {Silk}},\ }\href
  {\doibase 10.1016/j.physrep.2004.08.031} {\bibfield  {journal} {\bibinfo
  {journal} {Phys. Rept.}\ }\textbf {\bibinfo {volume} {405}},\ \bibinfo
  {pages} {279} (\bibinfo {year} {2005})},\ \Eprint
  {http://arxiv.org/abs/hep-ph/0404175} {arXiv:hep-ph/0404175 [hep-ph]}
  \BibitemShut {NoStop}%
\bibitem [{\citenamefont {Bergstrom}(2012)}]{Bergstrom:2012fi}%
  \BibitemOpen
  \bibfield  {author} {\bibinfo {author} {\bibfnamefont {L.}~\bibnamefont
  {Bergstrom}},\ }\href {\doibase 10.1002/andp.201200116} {\bibfield  {journal}
  {\bibinfo  {journal} {Annalen Phys.}\ }\textbf {\bibinfo {volume} {524}},\
  \bibinfo {pages} {479} (\bibinfo {year} {2012})},\ \Eprint
  {http://arxiv.org/abs/1205.4882} {arXiv:1205.4882 [astro-ph.HE]} \BibitemShut
  {NoStop}%
\bibitem [{\citenamefont {Kusenko}\ and\ \citenamefont
  {Rosenberg}(2013)}]{Kusenko:2013saa}%
  \BibitemOpen
  \bibfield  {author} {\bibinfo {author} {\bibfnamefont {A.}~\bibnamefont
  {Kusenko}}\ and\ \bibinfo {author} {\bibfnamefont {L.~J.}\ \bibnamefont
  {Rosenberg}},\ }in\ \href@noop {} {\emph {\bibinfo {booktitle} {Proceedings,
  Community Summer Study 2013: Snowmass on the Mississippi (CSS2013):
  Minneapolis, MN, USA, July 29-August 6, 2013}}}\ (\bibinfo {year} {2013})\
  \Eprint {http://arxiv.org/abs/1310.8642} {arXiv:1310.8642 [hep-ph]}
  \BibitemShut {NoStop}%
\bibitem [{\citenamefont {Klypin}\ \emph {et~al.}(1999)\citenamefont {Klypin},
  \citenamefont {Kravtsov}, \citenamefont {Valenzuela},\ and\ \citenamefont
  {Prada}}]{Klypin:1999uc}%
  \BibitemOpen
  \bibfield  {author} {\bibinfo {author} {\bibfnamefont {A.~A.}\ \bibnamefont
  {Klypin}}, \bibinfo {author} {\bibfnamefont {A.~V.}\ \bibnamefont
  {Kravtsov}}, \bibinfo {author} {\bibfnamefont {O.}~\bibnamefont
  {Valenzuela}}, \ and\ \bibinfo {author} {\bibfnamefont {F.}~\bibnamefont
  {Prada}},\ }\href {\doibase 10.1086/307643} {\bibfield  {journal} {\bibinfo
  {journal} {Astrophys. J.}\ }\textbf {\bibinfo {volume} {522}},\ \bibinfo
  {pages} {82} (\bibinfo {year} {1999})},\ \Eprint
  {http://arxiv.org/abs/astro-ph/9901240} {arXiv:astro-ph/9901240 [astro-ph]}
  \BibitemShut {NoStop}%
\bibitem [{\citenamefont {Moore}\ \emph
  {et~al.}(1999{\natexlab{a}})\citenamefont {Moore}, \citenamefont {Ghigna},
  \citenamefont {Governato}, \citenamefont {Lake}, \citenamefont {Quinn},
  \citenamefont {Stadel},\ and\ \citenamefont {Tozzi}}]{Moore:1999nt}%
  \BibitemOpen
  \bibfield  {author} {\bibinfo {author} {\bibfnamefont {B.}~\bibnamefont
  {Moore}}, \bibinfo {author} {\bibfnamefont {S.}~\bibnamefont {Ghigna}},
  \bibinfo {author} {\bibfnamefont {F.}~\bibnamefont {Governato}}, \bibinfo
  {author} {\bibfnamefont {G.}~\bibnamefont {Lake}}, \bibinfo {author}
  {\bibfnamefont {T.~R.}\ \bibnamefont {Quinn}}, \bibinfo {author}
  {\bibfnamefont {J.}~\bibnamefont {Stadel}}, \ and\ \bibinfo {author}
  {\bibfnamefont {P.}~\bibnamefont {Tozzi}},\ }\href {\doibase 10.1086/312287}
  {\bibfield  {journal} {\bibinfo  {journal} {Astrophys. J.}\ }\textbf
  {\bibinfo {volume} {524}},\ \bibinfo {pages} {L19} (\bibinfo {year}
  {1999}{\natexlab{a}})},\ \Eprint {http://arxiv.org/abs/astro-ph/9907411}
  {arXiv:astro-ph/9907411 [astro-ph]} \BibitemShut {NoStop}%
\bibitem [{\citenamefont {Boylan-Kolchin}\ \emph {et~al.}(2012)\citenamefont
  {Boylan-Kolchin}, \citenamefont {Bullock},\ and\ \citenamefont
  {Kaplinghat}}]{BoylanKolchin:2011dk}%
  \BibitemOpen
  \bibfield  {author} {\bibinfo {author} {\bibfnamefont {M.}~\bibnamefont
  {Boylan-Kolchin}}, \bibinfo {author} {\bibfnamefont {J.~S.}\ \bibnamefont
  {Bullock}}, \ and\ \bibinfo {author} {\bibfnamefont {M.}~\bibnamefont
  {Kaplinghat}},\ }\href {\doibase 10.1111/j.1365-2966.2012.20695.x} {\bibfield
   {journal} {\bibinfo  {journal} {Mon. Not. Roy. Astron. Soc.}\ }\textbf
  {\bibinfo {volume} {422}},\ \bibinfo {pages} {1203} (\bibinfo {year}
  {2012})},\ \Eprint {http://arxiv.org/abs/1111.2048} {arXiv:1111.2048
  [astro-ph.CO]} \BibitemShut {NoStop}%
\bibitem [{\citenamefont {Wang}\ \emph {et~al.}(2016)\citenamefont {Wang},
  \citenamefont {Gonzalez-Perez}, \citenamefont {Xie}, \citenamefont {Cooper},
  \citenamefont {Frenk}, \citenamefont {Gao}, \citenamefont {Hellwing},
  \citenamefont {Helly}, \citenamefont {Lovell},\ and\ \citenamefont
  {Jiang}}]{Wang:2016rio}%
  \BibitemOpen
  \bibfield  {author} {\bibinfo {author} {\bibfnamefont {L.}~\bibnamefont
  {Wang}}, \bibinfo {author} {\bibfnamefont {V.}~\bibnamefont
  {Gonzalez-Perez}}, \bibinfo {author} {\bibfnamefont {L.}~\bibnamefont {Xie}},
  \bibinfo {author} {\bibfnamefont {A.~P.}\ \bibnamefont {Cooper}}, \bibinfo
  {author} {\bibfnamefont {C.~S.}\ \bibnamefont {Frenk}}, \bibinfo {author}
  {\bibfnamefont {L.}~\bibnamefont {Gao}}, \bibinfo {author} {\bibfnamefont
  {W.~A.}\ \bibnamefont {Hellwing}}, \bibinfo {author} {\bibfnamefont
  {J.}~\bibnamefont {Helly}}, \bibinfo {author} {\bibfnamefont {M.~R.}\
  \bibnamefont {Lovell}}, \ and\ \bibinfo {author} {\bibfnamefont
  {L.}~\bibnamefont {Jiang}},\ }\href@noop {} {\  (\bibinfo {year} {2016})},\
  \Eprint {http://arxiv.org/abs/1612.04540} {arXiv:1612.04540 [astro-ph.GA]}
  \BibitemShut {NoStop}%
\bibitem [{\citenamefont {Lovell}\ \emph
  {et~al.}(2016{\natexlab{a}})\citenamefont {Lovell}, \citenamefont
  {Gonzalez-Perez}, \citenamefont {Bose}, \citenamefont {Boyarsky},
  \citenamefont {Cole}, \citenamefont {Frenk},\ and\ \citenamefont
  {Ruchayskiy}}]{Lovell:2016nkp}%
  \BibitemOpen
  \bibfield  {author} {\bibinfo {author} {\bibfnamefont {M.~R.}\ \bibnamefont
  {Lovell}}, \bibinfo {author} {\bibfnamefont {V.}~\bibnamefont
  {Gonzalez-Perez}}, \bibinfo {author} {\bibfnamefont {S.}~\bibnamefont
  {Bose}}, \bibinfo {author} {\bibfnamefont {A.}~\bibnamefont {Boyarsky}},
  \bibinfo {author} {\bibfnamefont {S.}~\bibnamefont {Cole}}, \bibinfo {author}
  {\bibfnamefont {C.~S.}\ \bibnamefont {Frenk}}, \ and\ \bibinfo {author}
  {\bibfnamefont {O.}~\bibnamefont {Ruchayskiy}},\ }\href@noop {} {\  (\bibinfo
  {year} {2016}{\natexlab{a}})},\ \Eprint {http://arxiv.org/abs/1611.00005}
  {arXiv:1611.00005 [astro-ph.GA]} \BibitemShut {NoStop}%
\bibitem [{\citenamefont {Sawala}\ \emph {et~al.}(2013)\citenamefont {Sawala},
  \citenamefont {Frenk}, \citenamefont {Crain}, \citenamefont {Jenkins},
  \citenamefont {Schaye}, \citenamefont {Theuns},\ and\ \citenamefont
  {Zavala}}]{Sawala:2012cn}%
  \BibitemOpen
  \bibfield  {author} {\bibinfo {author} {\bibfnamefont {T.}~\bibnamefont
  {Sawala}}, \bibinfo {author} {\bibfnamefont {C.~S.}\ \bibnamefont {Frenk}},
  \bibinfo {author} {\bibfnamefont {R.~A.}\ \bibnamefont {Crain}}, \bibinfo
  {author} {\bibfnamefont {A.}~\bibnamefont {Jenkins}}, \bibinfo {author}
  {\bibfnamefont {J.}~\bibnamefont {Schaye}}, \bibinfo {author} {\bibfnamefont
  {T.}~\bibnamefont {Theuns}}, \ and\ \bibinfo {author} {\bibfnamefont
  {J.}~\bibnamefont {Zavala}},\ }\href {\doibase 10.1093/mnras/stt259}
  {\bibfield  {journal} {\bibinfo  {journal} {Mon. Not. Roy. Astron. Soc.}\
  }\textbf {\bibinfo {volume} {431}},\ \bibinfo {pages} {1366} (\bibinfo {year}
  {2013})},\ \Eprint {http://arxiv.org/abs/1206.6495} {arXiv:1206.6495
  [astro-ph.CO]} \BibitemShut {NoStop}%
\bibitem [{\citenamefont {Sawala}\ \emph {et~al.}(2016)\citenamefont {Sawala}
  \emph {et~al.}}]{Sawala:2015cdf}%
  \BibitemOpen
  \bibfield  {author} {\bibinfo {author} {\bibfnamefont {T.}~\bibnamefont
  {Sawala}} \emph {et~al.},\ }\href {\doibase 10.1093/mnras/stw145} {\bibfield
  {journal} {\bibinfo  {journal} {Mon. Not. Roy. Astron. Soc.}\ }\textbf
  {\bibinfo {volume} {457}},\ \bibinfo {pages} {1931} (\bibinfo {year}
  {2016})},\ \Eprint {http://arxiv.org/abs/1511.01098} {arXiv:1511.01098
  [astro-ph.GA]} \BibitemShut {NoStop}%
\bibitem [{\citenamefont {Fattahi}\ \emph {et~al.}(2016)\citenamefont
  {Fattahi}, \citenamefont {Navarro}, \citenamefont {Sawala}, \citenamefont
  {Frenk}, \citenamefont {Sales}, \citenamefont {Oman}, \citenamefont
  {Schaller},\ and\ \citenamefont {Wang}}]{Fattahi:2016nld}%
  \BibitemOpen
  \bibfield  {author} {\bibinfo {author} {\bibfnamefont {A.}~\bibnamefont
  {Fattahi}}, \bibinfo {author} {\bibfnamefont {J.~F.}\ \bibnamefont
  {Navarro}}, \bibinfo {author} {\bibfnamefont {T.}~\bibnamefont {Sawala}},
  \bibinfo {author} {\bibfnamefont {C.~S.}\ \bibnamefont {Frenk}}, \bibinfo
  {author} {\bibfnamefont {L.~V.}\ \bibnamefont {Sales}}, \bibinfo {author}
  {\bibfnamefont {K.}~\bibnamefont {Oman}}, \bibinfo {author} {\bibfnamefont
  {M.}~\bibnamefont {Schaller}}, \ and\ \bibinfo {author} {\bibfnamefont
  {J.}~\bibnamefont {Wang}},\ }\href@noop {} {\  (\bibinfo {year} {2016})},\
  \Eprint {http://arxiv.org/abs/1607.06479} {arXiv:1607.06479 [astro-ph.GA]}
  \BibitemShut {NoStop}%
\bibitem [{\citenamefont {Polisensky}\ and\ \citenamefont
  {Ricotti}(2014)}]{Polisensky:2013ppa}%
  \BibitemOpen
  \bibfield  {author} {\bibinfo {author} {\bibfnamefont {E.}~\bibnamefont
  {Polisensky}}\ and\ \bibinfo {author} {\bibfnamefont {M.}~\bibnamefont
  {Ricotti}},\ }\href {\doibase 10.1093/mnras/stt2105} {\bibfield  {journal}
  {\bibinfo  {journal} {Mon. Not. Roy. Astron. Soc.}\ }\textbf {\bibinfo
  {volume} {437}},\ \bibinfo {pages} {2922} (\bibinfo {year} {2014})},\ \Eprint
  {http://arxiv.org/abs/1310.0430} {arXiv:1310.0430 [astro-ph.CO]} \BibitemShut
  {NoStop}%
\bibitem [{\citenamefont {Vogelsberger}\ \emph {et~al.}(2012)\citenamefont
  {Vogelsberger}, \citenamefont {Zavala},\ and\ \citenamefont
  {Loeb}}]{Vogelsberger:2012ku}%
  \BibitemOpen
  \bibfield  {author} {\bibinfo {author} {\bibfnamefont {M.}~\bibnamefont
  {Vogelsberger}}, \bibinfo {author} {\bibfnamefont {J.}~\bibnamefont
  {Zavala}}, \ and\ \bibinfo {author} {\bibfnamefont {A.}~\bibnamefont
  {Loeb}},\ }\href {\doibase 10.1111/j.1365-2966.2012.21182.x} {\bibfield
  {journal} {\bibinfo  {journal} {Mon. Not. Roy. Astron. Soc.}\ }\textbf
  {\bibinfo {volume} {423}},\ \bibinfo {pages} {3740} (\bibinfo {year}
  {2012})},\ \Eprint {http://arxiv.org/abs/1201.5892} {arXiv:1201.5892
  [astro-ph.CO]} \BibitemShut {NoStop}%
\bibitem [{\citenamefont {Schewtschenko}\ \emph {et~al.}(2016)\citenamefont
  {Schewtschenko}, \citenamefont {Baugh}, \citenamefont {Wilkinson},
  \citenamefont {Boehm}, \citenamefont {Pascoli},\ and\ \citenamefont
  {Sawala}}]{Schewtschenko:2015rno}%
  \BibitemOpen
  \bibfield  {author} {\bibinfo {author} {\bibfnamefont {J.~A.}\ \bibnamefont
  {Schewtschenko}}, \bibinfo {author} {\bibfnamefont {C.~M.}\ \bibnamefont
  {Baugh}}, \bibinfo {author} {\bibfnamefont {R.~J.}\ \bibnamefont
  {Wilkinson}}, \bibinfo {author} {\bibfnamefont {C.}~\bibnamefont {Boehm}},
  \bibinfo {author} {\bibfnamefont {S.}~\bibnamefont {Pascoli}}, \ and\
  \bibinfo {author} {\bibfnamefont {T.}~\bibnamefont {Sawala}},\ }\href
  {\doibase 10.1093/mnras/stw1078} {\bibfield  {journal} {\bibinfo  {journal}
  {Mon. Not. Roy. Astron. Soc.}\ }\textbf {\bibinfo {volume} {461}},\ \bibinfo
  {pages} {2282} (\bibinfo {year} {2016})},\ \Eprint
  {http://arxiv.org/abs/1512.06774} {arXiv:1512.06774 [astro-ph.CO]}
  \BibitemShut {NoStop}%
\bibitem [{\citenamefont {Lovell}\ \emph {et~al.}(2012)\citenamefont {Lovell},
  \citenamefont {Eke}, \citenamefont {Frenk}, \citenamefont {Gao},
  \citenamefont {Jenkins}, \citenamefont {Theuns}, \citenamefont {Wang},
  \citenamefont {White}, \citenamefont {Boyarsky},\ and\ \citenamefont
  {Ruchayskiy}}]{Lovell:2011rd}%
  \BibitemOpen
  \bibfield  {author} {\bibinfo {author} {\bibfnamefont {M.~R.}\ \bibnamefont
  {Lovell}}, \bibinfo {author} {\bibfnamefont {V.}~\bibnamefont {Eke}},
  \bibinfo {author} {\bibfnamefont {C.~S.}\ \bibnamefont {Frenk}}, \bibinfo
  {author} {\bibfnamefont {L.}~\bibnamefont {Gao}}, \bibinfo {author}
  {\bibfnamefont {A.}~\bibnamefont {Jenkins}}, \bibinfo {author} {\bibfnamefont
  {T.}~\bibnamefont {Theuns}}, \bibinfo {author} {\bibfnamefont
  {J.}~\bibnamefont {Wang}}, \bibinfo {author} {\bibfnamefont {D.~M.}\
  \bibnamefont {White}}, \bibinfo {author} {\bibfnamefont {A.}~\bibnamefont
  {Boyarsky}}, \ and\ \bibinfo {author} {\bibfnamefont {O.}~\bibnamefont
  {Ruchayskiy}},\ }\href {\doibase 10.1111/j.1365-2966.2011.20200.x} {\bibfield
   {journal} {\bibinfo  {journal} {Mon. Not. Roy. Astron. Soc.}\ }\textbf
  {\bibinfo {volume} {420}},\ \bibinfo {pages} {2318} (\bibinfo {year}
  {2012})},\ \Eprint {http://arxiv.org/abs/1104.2929} {arXiv:1104.2929
  [astro-ph.CO]} \BibitemShut {NoStop}%
\bibitem [{\citenamefont {Lovell}\ \emph {et~al.}(2014)\citenamefont {Lovell},
  \citenamefont {Frenk}, \citenamefont {Eke}, \citenamefont {Jenkins},
  \citenamefont {Gao},\ and\ \citenamefont {Theuns}}]{Lovell:2013ola}%
  \BibitemOpen
  \bibfield  {author} {\bibinfo {author} {\bibfnamefont {M.~R.}\ \bibnamefont
  {Lovell}}, \bibinfo {author} {\bibfnamefont {C.~S.}\ \bibnamefont {Frenk}},
  \bibinfo {author} {\bibfnamefont {V.~R.}\ \bibnamefont {Eke}}, \bibinfo
  {author} {\bibfnamefont {A.}~\bibnamefont {Jenkins}}, \bibinfo {author}
  {\bibfnamefont {L.}~\bibnamefont {Gao}}, \ and\ \bibinfo {author}
  {\bibfnamefont {T.}~\bibnamefont {Theuns}},\ }\href {\doibase
  10.1093/mnras/stt2431} {\bibfield  {journal} {\bibinfo  {journal} {Mon. Not.
  Roy. Astron. Soc.}\ }\textbf {\bibinfo {volume} {439}},\ \bibinfo {pages}
  {300} (\bibinfo {year} {2014})},\ \Eprint {http://arxiv.org/abs/1308.1399}
  {arXiv:1308.1399 [astro-ph.CO]} \BibitemShut {NoStop}%
\bibitem [{\citenamefont {Lovell}\ \emph
  {et~al.}(2016{\natexlab{b}})\citenamefont {Lovell}, \citenamefont {Bose},
  \citenamefont {Boyarsky}, \citenamefont {Cole}, \citenamefont {Frenk},
  \citenamefont {Gonzalez-Perez}, \citenamefont {Kennedy}, \citenamefont
  {Ruchayskiy},\ and\ \citenamefont {Smith}}]{Lovell:2015psz}%
  \BibitemOpen
  \bibfield  {author} {\bibinfo {author} {\bibfnamefont {M.~R.}\ \bibnamefont
  {Lovell}}, \bibinfo {author} {\bibfnamefont {S.}~\bibnamefont {Bose}},
  \bibinfo {author} {\bibfnamefont {A.}~\bibnamefont {Boyarsky}}, \bibinfo
  {author} {\bibfnamefont {S.}~\bibnamefont {Cole}}, \bibinfo {author}
  {\bibfnamefont {C.~S.}\ \bibnamefont {Frenk}}, \bibinfo {author}
  {\bibfnamefont {V.}~\bibnamefont {Gonzalez-Perez}}, \bibinfo {author}
  {\bibfnamefont {R.}~\bibnamefont {Kennedy}}, \bibinfo {author} {\bibfnamefont
  {O.}~\bibnamefont {Ruchayskiy}}, \ and\ \bibinfo {author} {\bibfnamefont
  {A.}~\bibnamefont {Smith}},\ }\href {\doibase 10.1093/mnras/stw1317}
  {\bibfield  {journal} {\bibinfo  {journal} {Mon. Not. Roy. Astron. Soc.}\
  }\textbf {\bibinfo {volume} {461}},\ \bibinfo {pages} {60} (\bibinfo {year}
  {2016}{\natexlab{b}})},\ \Eprint {http://arxiv.org/abs/1511.04078}
  {arXiv:1511.04078 [astro-ph.CO]} \BibitemShut {NoStop}%
\bibitem [{\citenamefont {Nakama}\ \emph {et~al.}(2017)\citenamefont {Nakama},
  \citenamefont {Chluba},\ and\ \citenamefont {Kamionkowski}}]{Nakama:2017ohe}%
  \BibitemOpen
  \bibfield  {author} {\bibinfo {author} {\bibfnamefont {T.}~\bibnamefont
  {Nakama}}, \bibinfo {author} {\bibfnamefont {J.}~\bibnamefont {Chluba}}, \
  and\ \bibinfo {author} {\bibfnamefont {M.}~\bibnamefont {Kamionkowski}},\
  }\href@noop {} {\  (\bibinfo {year} {2017})},\ \Eprint
  {http://arxiv.org/abs/1703.10559} {arXiv:1703.10559 [astro-ph.CO]}
  \BibitemShut {NoStop}%
\bibitem [{\citenamefont {Hildebrandt}\ \emph {et~al.}(2016)\citenamefont
  {Hildebrandt} \emph {et~al.}}]{Hildebrandt:2016iqg}%
  \BibitemOpen
  \bibfield  {author} {\bibinfo {author} {\bibfnamefont {H.}~\bibnamefont
  {Hildebrandt}} \emph {et~al.},\ }\href@noop {} {\  (\bibinfo {year}
  {2016})},\ \Eprint {http://arxiv.org/abs/1606.05338} {arXiv:1606.05338
  [astro-ph.CO]} \BibitemShut {NoStop}%
\bibitem [{\citenamefont {Joudaki}\ \emph
  {et~al.}(2016{\natexlab{a}})\citenamefont {Joudaki} \emph
  {et~al.}}]{Joudaki:2016kym}%
  \BibitemOpen
  \bibfield  {author} {\bibinfo {author} {\bibfnamefont {S.}~\bibnamefont
  {Joudaki}} \emph {et~al.},\ }\href@noop {} {\  (\bibinfo {year}
  {2016}{\natexlab{a}})},\ \Eprint {http://arxiv.org/abs/1610.04606}
  {arXiv:1610.04606 [astro-ph.CO]} \BibitemShut {NoStop}%
\bibitem [{\citenamefont {Aghanim}\ \emph {et~al.}(2016)\citenamefont {Aghanim}
  \emph {et~al.}}]{Aghanim:2015xee}%
  \BibitemOpen
  \bibfield  {author} {\bibinfo {author} {\bibfnamefont {N.}~\bibnamefont
  {Aghanim}} \emph {et~al.} (\bibinfo {collaboration} {Planck}),\ }\href
  {\doibase 10.1051/0004-6361/201526926} {\bibfield  {journal} {\bibinfo
  {journal} {Astron. Astrophys.}\ }\textbf {\bibinfo {volume} {594}},\ \bibinfo
  {pages} {A11} (\bibinfo {year} {2016})},\ \Eprint
  {http://arxiv.org/abs/1507.02704} {arXiv:1507.02704 [astro-ph.CO]}
  \BibitemShut {NoStop}%
\bibitem [{\citenamefont {Kilbinger}\ \emph {et~al.}(2013)\citenamefont
  {Kilbinger} \emph {et~al.}}]{Kilbinger:2012qz}%
  \BibitemOpen
  \bibfield  {author} {\bibinfo {author} {\bibfnamefont {M.}~\bibnamefont
  {Kilbinger}} \emph {et~al.},\ }\href {\doibase 10.1093/mnras/stt041}
  {\bibfield  {journal} {\bibinfo  {journal} {Mon. Not. Roy. Astron. Soc.}\
  }\textbf {\bibinfo {volume} {430}},\ \bibinfo {pages} {2200} (\bibinfo {year}
  {2013})},\ \Eprint {http://arxiv.org/abs/1212.3338} {arXiv:1212.3338
  [astro-ph.CO]} \BibitemShut {NoStop}%
\bibitem [{\citenamefont {Heymans}\ \emph {et~al.}(2013)\citenamefont {Heymans}
  \emph {et~al.}}]{Heymans:2013fya}%
  \BibitemOpen
  \bibfield  {author} {\bibinfo {author} {\bibfnamefont {C.}~\bibnamefont
  {Heymans}} \emph {et~al.},\ }\href {\doibase 10.1093/mnras/stt601} {\bibfield
   {journal} {\bibinfo  {journal} {Mon. Not. Roy. Astron. Soc.}\ }\textbf
  {\bibinfo {volume} {432}},\ \bibinfo {pages} {2433} (\bibinfo {year}
  {2013})},\ \Eprint {http://arxiv.org/abs/1303.1808} {arXiv:1303.1808
  [astro-ph.CO]} \BibitemShut {NoStop}%
\bibitem [{\citenamefont {Lemos}\ \emph {et~al.}(2017)\citenamefont {Lemos},
  \citenamefont {Challinor},\ and\ \citenamefont {Efstathiou}}]{Lemos:2017arq}%
  \BibitemOpen
  \bibfield  {author} {\bibinfo {author} {\bibfnamefont {P.}~\bibnamefont
  {Lemos}}, \bibinfo {author} {\bibfnamefont {A.}~\bibnamefont {Challinor}}, \
  and\ \bibinfo {author} {\bibfnamefont {G.}~\bibnamefont {Efstathiou}}\
  }(\bibinfo {year} {2017})\ \Eprint {http://arxiv.org/abs/1704.01054}
  {arXiv:1704.01054 [astro-ph.CO]} \BibitemShut {NoStop}%
\bibitem [{\citenamefont {Kilbinger}\ \emph {et~al.}(2017)\citenamefont
  {Kilbinger} \emph {et~al.}}]{Kilbinger:2017lvu}%
  \BibitemOpen
  \bibfield  {author} {\bibinfo {author} {\bibfnamefont {M.}~\bibnamefont
  {Kilbinger}} \emph {et~al.},\ }\href@noop {} {\  (\bibinfo {year} {2017})},\
  \Eprint {http://arxiv.org/abs/1702.05301} {arXiv:1702.05301 [astro-ph.CO]}
  \BibitemShut {NoStop}%
\bibitem [{\citenamefont {Kitching}\ \emph
  {et~al.}(2016{\natexlab{a}})\citenamefont {Kitching}, \citenamefont {Alsing},
  \citenamefont {Heavens}, \citenamefont {Jimenez}, \citenamefont {McEwen},\
  and\ \citenamefont {Verde}}]{Kitching:2016zkn}%
  \BibitemOpen
  \bibfield  {author} {\bibinfo {author} {\bibfnamefont {T.~D.}\ \bibnamefont
  {Kitching}}, \bibinfo {author} {\bibfnamefont {J.}~\bibnamefont {Alsing}},
  \bibinfo {author} {\bibfnamefont {A.~F.}\ \bibnamefont {Heavens}}, \bibinfo
  {author} {\bibfnamefont {R.}~\bibnamefont {Jimenez}}, \bibinfo {author}
  {\bibfnamefont {J.~D.}\ \bibnamefont {McEwen}}, \ and\ \bibinfo {author}
  {\bibfnamefont {L.}~\bibnamefont {Verde}},\ }\href@noop {} {\  (\bibinfo
  {year} {2016}{\natexlab{a}})},\ \Eprint {http://arxiv.org/abs/1611.04954}
  {arXiv:1611.04954 [astro-ph.CO]} \BibitemShut {NoStop}%
\bibitem [{\citenamefont {Palazzo}\ \emph {et~al.}(2007)\citenamefont
  {Palazzo}, \citenamefont {Cumberbatch}, \citenamefont {Slosar},\ and\
  \citenamefont {Silk}}]{Palazzo:2007gz}%
  \BibitemOpen
  \bibfield  {author} {\bibinfo {author} {\bibfnamefont {A.}~\bibnamefont
  {Palazzo}}, \bibinfo {author} {\bibfnamefont {D.}~\bibnamefont
  {Cumberbatch}}, \bibinfo {author} {\bibfnamefont {A.}~\bibnamefont {Slosar}},
  \ and\ \bibinfo {author} {\bibfnamefont {J.}~\bibnamefont {Silk}},\ }\href
  {\doibase 10.1103/PhysRevD.76.103511} {\bibfield  {journal} {\bibinfo
  {journal} {Phys. Rev.}\ }\textbf {\bibinfo {volume} {D76}},\ \bibinfo {pages}
  {103511} (\bibinfo {year} {2007})},\ \Eprint {http://arxiv.org/abs/0707.1495}
  {arXiv:0707.1495 [astro-ph]} \BibitemShut {NoStop}%
\bibitem [{\citenamefont {Boyarsky}\ \emph {et~al.}(2009)\citenamefont
  {Boyarsky}, \citenamefont {Lesgourgues}, \citenamefont {Ruchayskiy},\ and\
  \citenamefont {Viel}}]{Boyarsky:2008xj}%
  \BibitemOpen
  \bibfield  {author} {\bibinfo {author} {\bibfnamefont {A.}~\bibnamefont
  {Boyarsky}}, \bibinfo {author} {\bibfnamefont {J.}~\bibnamefont
  {Lesgourgues}}, \bibinfo {author} {\bibfnamefont {O.}~\bibnamefont
  {Ruchayskiy}}, \ and\ \bibinfo {author} {\bibfnamefont {M.}~\bibnamefont
  {Viel}},\ }\href {\doibase 10.1088/1475-7516/2009/05/012} {\bibfield
  {journal} {\bibinfo  {journal} {JCAP}\ }\textbf {\bibinfo {volume} {0905}},\
  \bibinfo {pages} {012} (\bibinfo {year} {2009})},\ \Eprint
  {http://arxiv.org/abs/0812.0010} {arXiv:0812.0010 [astro-ph]} \BibitemShut
  {NoStop}%
\bibitem [{\citenamefont {Maccio}\ \emph {et~al.}(2013)\citenamefont {Maccio},
  \citenamefont {Ruchayskiy}, \citenamefont {Boyarsky},\ and\ \citenamefont
  {Munoz-Cuartas}}]{Maccio:2012rjx}%
  \BibitemOpen
  \bibfield  {author} {\bibinfo {author} {\bibfnamefont {A.~V.}\ \bibnamefont
  {Maccio}}, \bibinfo {author} {\bibfnamefont {O.}~\bibnamefont {Ruchayskiy}},
  \bibinfo {author} {\bibfnamefont {A.}~\bibnamefont {Boyarsky}}, \ and\
  \bibinfo {author} {\bibfnamefont {J.~C.}\ \bibnamefont {Munoz-Cuartas}},\
  }\href {\doibase 10.1093/mnras/sts078} {\bibfield  {journal} {\bibinfo
  {journal} {Mon. Not. Roy. Astron. Soc.}\ }\textbf {\bibinfo {volume} {428}},\
  \bibinfo {pages} {882} (\bibinfo {year} {2013})},\ \Eprint
  {http://arxiv.org/abs/1202.2858} {arXiv:1202.2858 [astro-ph.CO]} \BibitemShut
  {NoStop}%
\bibitem [{\citenamefont {Anderhalden}\ \emph {et~al.}(2012)\citenamefont
  {Anderhalden}, \citenamefont {Diemand}, \citenamefont {Bertone},
  \citenamefont {Maccio},\ and\ \citenamefont
  {Schneider}}]{Anderhalden:2012qt}%
  \BibitemOpen
  \bibfield  {author} {\bibinfo {author} {\bibfnamefont {D.}~\bibnamefont
  {Anderhalden}}, \bibinfo {author} {\bibfnamefont {J.}~\bibnamefont
  {Diemand}}, \bibinfo {author} {\bibfnamefont {G.}~\bibnamefont {Bertone}},
  \bibinfo {author} {\bibfnamefont {A.~V.}\ \bibnamefont {Maccio}}, \ and\
  \bibinfo {author} {\bibfnamefont {A.}~\bibnamefont {Schneider}},\ }\href
  {\doibase 10.1088/1475-7516/2012/10/047} {\bibfield  {journal} {\bibinfo
  {journal} {JCAP}\ }\textbf {\bibinfo {volume} {1210}},\ \bibinfo {pages}
  {047} (\bibinfo {year} {2012})},\ \Eprint {http://arxiv.org/abs/1206.3788}
  {arXiv:1206.3788 [astro-ph.CO]} \BibitemShut {NoStop}%
\bibitem [{\citenamefont {Anderhalden}\ \emph {et~al.}(2013)\citenamefont
  {Anderhalden}, \citenamefont {Schneider}, \citenamefont {Maccio},
  \citenamefont {Diemand},\ and\ \citenamefont {Bertone}}]{Anderhalden:2012jc}%
  \BibitemOpen
  \bibfield  {author} {\bibinfo {author} {\bibfnamefont {D.}~\bibnamefont
  {Anderhalden}}, \bibinfo {author} {\bibfnamefont {A.}~\bibnamefont
  {Schneider}}, \bibinfo {author} {\bibfnamefont {A.~V.}\ \bibnamefont
  {Maccio}}, \bibinfo {author} {\bibfnamefont {J.}~\bibnamefont {Diemand}}, \
  and\ \bibinfo {author} {\bibfnamefont {G.}~\bibnamefont {Bertone}},\ }\href
  {\doibase 10.1088/1475-7516/2013/03/014} {\bibfield  {journal} {\bibinfo
  {journal} {JCAP}\ }\textbf {\bibinfo {volume} {1303}},\ \bibinfo {pages}
  {014} (\bibinfo {year} {2013})},\ \Eprint {http://arxiv.org/abs/1212.2967}
  {arXiv:1212.2967 [astro-ph.CO]} \BibitemShut {NoStop}%
\bibitem [{\citenamefont {Diamanti}\ \emph {et~al.}(2017)\citenamefont
  {Diamanti}, \citenamefont {Ando}, \citenamefont {Gariazzo}, \citenamefont
  {Mena},\ and\ \citenamefont {Weniger}}]{Diamanti:2017xfo}%
  \BibitemOpen
  \bibfield  {author} {\bibinfo {author} {\bibfnamefont {R.}~\bibnamefont
  {Diamanti}}, \bibinfo {author} {\bibfnamefont {S.}~\bibnamefont {Ando}},
  \bibinfo {author} {\bibfnamefont {S.}~\bibnamefont {Gariazzo}}, \bibinfo
  {author} {\bibfnamefont {O.}~\bibnamefont {Mena}}, \ and\ \bibinfo {author}
  {\bibfnamefont {C.}~\bibnamefont {Weniger}},\ }\href@noop {} {\  (\bibinfo
  {year} {2017})},\ \Eprint {http://arxiv.org/abs/1701.03128} {arXiv:1701.03128
  [astro-ph.CO]} \BibitemShut {NoStop}%
\bibitem [{\citenamefont {Weinberg}\ \emph {et~al.}(2014)\citenamefont
  {Weinberg}, \citenamefont {Bullock}, \citenamefont {Governato}, \citenamefont
  {Kuzio~de Naray},\ and\ \citenamefont {Peter}}]{Weinberg:2013aya}%
  \BibitemOpen
  \bibfield  {author} {\bibinfo {author} {\bibfnamefont {D.~H.}\ \bibnamefont
  {Weinberg}}, \bibinfo {author} {\bibfnamefont {J.~S.}\ \bibnamefont
  {Bullock}}, \bibinfo {author} {\bibfnamefont {F.}~\bibnamefont {Governato}},
  \bibinfo {author} {\bibfnamefont {R.}~\bibnamefont {Kuzio~de Naray}}, \ and\
  \bibinfo {author} {\bibfnamefont {A.~H.~G.}\ \bibnamefont {Peter}},\
  }\bibfield  {booktitle} {\emph {\bibinfo {booktitle} {Sackler Colloquium:
  Dark Matter Universe: On the Threshhold of Discovery Irvine, USA, October
  18-20, 2012}},\ }\href {\doibase 10.1073/pnas.1308716112} {\bibfield
  {journal} {\bibinfo  {journal} {Proc. Nat. Acad. Sci.}\ }\textbf {\bibinfo
  {volume} {112}},\ \bibinfo {pages} {12249} (\bibinfo {year} {2014})},\
  \Eprint {http://arxiv.org/abs/1306.0913} {arXiv:1306.0913 [astro-ph.CO]}
  \BibitemShut {NoStop}%
\bibitem [{\citenamefont {Ali-Haimoud}\ and\ \citenamefont
  {Bird}(2012)}]{AliHaimoud:2012vj}%
  \BibitemOpen
  \bibfield  {author} {\bibinfo {author} {\bibfnamefont {Y.}~\bibnamefont
  {Ali-Haimoud}}\ and\ \bibinfo {author} {\bibfnamefont {S.}~\bibnamefont
  {Bird}},\ }\href {\doibase 10.1093/mnras/sts286} {\bibfield  {journal}
  {\bibinfo  {journal} {Mon. Not. Roy. Astron. Soc.}\ }\textbf {\bibinfo
  {volume} {428}},\ \bibinfo {pages} {3375} (\bibinfo {year} {2012})},\ \Eprint
  {http://arxiv.org/abs/1209.0461} {arXiv:1209.0461 [astro-ph.CO]} \BibitemShut
  {NoStop}%
\bibitem [{\citenamefont {Seljak}(2000)}]{Seljak:2000gq}%
  \BibitemOpen
  \bibfield  {author} {\bibinfo {author} {\bibfnamefont {U.}~\bibnamefont
  {Seljak}},\ }\href {\doibase 10.1046/j.1365-8711.2000.03715.x} {\bibfield
  {journal} {\bibinfo  {journal} {Mon. Not. Roy. Astron. Soc.}\ }\textbf
  {\bibinfo {volume} {318}},\ \bibinfo {pages} {203} (\bibinfo {year}
  {2000})},\ \Eprint {http://arxiv.org/abs/astro-ph/0001493}
  {arXiv:astro-ph/0001493 [astro-ph]} \BibitemShut {NoStop}%
\bibitem [{\citenamefont {Peacock}\ and\ \citenamefont
  {Smith}(2000)}]{Peacock:2000qk}%
  \BibitemOpen
  \bibfield  {author} {\bibinfo {author} {\bibfnamefont {J.~A.}\ \bibnamefont
  {Peacock}}\ and\ \bibinfo {author} {\bibfnamefont {R.~E.}\ \bibnamefont
  {Smith}},\ }\href {\doibase 10.1046/j.1365-8711.2000.03779.x} {\bibfield
  {journal} {\bibinfo  {journal} {Mon. Not. Roy. Astron. Soc.}\ }\textbf
  {\bibinfo {volume} {318}},\ \bibinfo {pages} {1144} (\bibinfo {year}
  {2000})},\ \Eprint {http://arxiv.org/abs/astro-ph/0005010}
  {arXiv:astro-ph/0005010 [astro-ph]} \BibitemShut {NoStop}%
\bibitem [{\citenamefont {Ma}\ and\ \citenamefont {Fry}(2000)}]{Ma:2000ik}%
  \BibitemOpen
  \bibfield  {author} {\bibinfo {author} {\bibfnamefont {C.-P.}\ \bibnamefont
  {Ma}}\ and\ \bibinfo {author} {\bibfnamefont {J.~N.}\ \bibnamefont {Fry}},\
  }\href {\doibase 10.1086/317146} {\bibfield  {journal} {\bibinfo  {journal}
  {Astrophys. J.}\ }\textbf {\bibinfo {volume} {543}},\ \bibinfo {pages} {503}
  (\bibinfo {year} {2000})},\ \Eprint {http://arxiv.org/abs/astro-ph/0003343}
  {arXiv:astro-ph/0003343 [astro-ph]} \BibitemShut {NoStop}%
\bibitem [{\citenamefont {Cooray}\ and\ \citenamefont
  {Sheth}(2002)}]{Cooray:2002dia}%
  \BibitemOpen
  \bibfield  {author} {\bibinfo {author} {\bibfnamefont {A.}~\bibnamefont
  {Cooray}}\ and\ \bibinfo {author} {\bibfnamefont {R.~K.}\ \bibnamefont
  {Sheth}},\ }\href {\doibase 10.1016/S0370-1573(02)00276-4} {\bibfield
  {journal} {\bibinfo  {journal} {Phys. Rept.}\ }\textbf {\bibinfo {volume}
  {372}},\ \bibinfo {pages} {1} (\bibinfo {year} {2002})},\ \Eprint
  {http://arxiv.org/abs/astro-ph/0206508} {arXiv:astro-ph/0206508 [astro-ph]}
  \BibitemShut {NoStop}%
\bibitem [{\citenamefont {Mead}\ \emph {et~al.}(2015)\citenamefont {Mead},
  \citenamefont {Peacock}, \citenamefont {Heymans}, \citenamefont {Joudaki},\
  and\ \citenamefont {Heavens}}]{Mead:2015yca}%
  \BibitemOpen
  \bibfield  {author} {\bibinfo {author} {\bibfnamefont {A.}~\bibnamefont
  {Mead}}, \bibinfo {author} {\bibfnamefont {J.}~\bibnamefont {Peacock}},
  \bibinfo {author} {\bibfnamefont {C.}~\bibnamefont {Heymans}}, \bibinfo
  {author} {\bibfnamefont {S.}~\bibnamefont {Joudaki}}, \ and\ \bibinfo
  {author} {\bibfnamefont {A.}~\bibnamefont {Heavens}},\ }\href {\doibase
  10.1093/mnras/stv2036} {\bibfield  {journal} {\bibinfo  {journal} {Mon. Not.
  Roy. Astron. Soc.}\ }\textbf {\bibinfo {volume} {454}},\ \bibinfo {pages}
  {1958} (\bibinfo {year} {2015})},\ \Eprint {http://arxiv.org/abs/1505.07833}
  {arXiv:1505.07833 [astro-ph.CO]} \BibitemShut {NoStop}%
\bibitem [{\citenamefont {Takahashi}\ \emph {et~al.}(2012)\citenamefont
  {Takahashi}, \citenamefont {Sato}, \citenamefont {Nishimichi}, \citenamefont
  {Taruya},\ and\ \citenamefont {Oguri}}]{Takahashi:2012em}%
  \BibitemOpen
  \bibfield  {author} {\bibinfo {author} {\bibfnamefont {R.}~\bibnamefont
  {Takahashi}}, \bibinfo {author} {\bibfnamefont {M.}~\bibnamefont {Sato}},
  \bibinfo {author} {\bibfnamefont {T.}~\bibnamefont {Nishimichi}}, \bibinfo
  {author} {\bibfnamefont {A.}~\bibnamefont {Taruya}}, \ and\ \bibinfo {author}
  {\bibfnamefont {M.}~\bibnamefont {Oguri}},\ }\href {\doibase
  10.1088/0004-637X/761/2/152} {\bibfield  {journal} {\bibinfo  {journal}
  {Astrophys. J.}\ }\textbf {\bibinfo {volume} {761}},\ \bibinfo {pages} {152}
  (\bibinfo {year} {2012})},\ \Eprint {http://arxiv.org/abs/1208.2701}
  {arXiv:1208.2701 [astro-ph.CO]} \BibitemShut {NoStop}%
\bibitem [{\citenamefont {Kamada}\ \emph {et~al.}(2016)\citenamefont {Kamada},
  \citenamefont {Inoue},\ and\ \citenamefont {Takahashi}}]{Kamada:2016vsc}%
  \BibitemOpen
  \bibfield  {author} {\bibinfo {author} {\bibfnamefont {A.}~\bibnamefont
  {Kamada}}, \bibinfo {author} {\bibfnamefont {K.~T.}\ \bibnamefont {Inoue}}, \
  and\ \bibinfo {author} {\bibfnamefont {T.}~\bibnamefont {Takahashi}},\ }\href
  {\doibase 10.1103/PhysRevD.94.023522} {\bibfield  {journal} {\bibinfo
  {journal} {Phys. Rev.}\ }\textbf {\bibinfo {volume} {D94}},\ \bibinfo {pages}
  {023522} (\bibinfo {year} {2016})},\ \Eprint
  {http://arxiv.org/abs/1604.01489} {arXiv:1604.01489 [astro-ph.CO]}
  \BibitemShut {NoStop}%
\bibitem [{\citenamefont {Takahashi}\ and\ \citenamefont
  {Inoue}(2014)}]{Takahashi:2013sna}%
  \BibitemOpen
  \bibfield  {author} {\bibinfo {author} {\bibfnamefont {R.}~\bibnamefont
  {Takahashi}}\ and\ \bibinfo {author} {\bibfnamefont {K.~T.}\ \bibnamefont
  {Inoue}},\ }\href {\doibase 10.1093/mnras/stu328} {\bibfield  {journal}
  {\bibinfo  {journal} {Mon. Not. Roy. Astron. Soc.}\ }\textbf {\bibinfo
  {volume} {440}},\ \bibinfo {pages} {870} (\bibinfo {year} {2014})},\ \Eprint
  {http://arxiv.org/abs/1308.4855} {arXiv:1308.4855 [astro-ph.CO]} \BibitemShut
  {NoStop}%
\bibitem [{\citenamefont {Inoue}\ \emph {et~al.}(2015)\citenamefont {Inoue},
  \citenamefont {Takahashi}, \citenamefont {Takahashi},\ and\ \citenamefont
  {Ishiyama}}]{Inoue:2014jka}%
  \BibitemOpen
  \bibfield  {author} {\bibinfo {author} {\bibfnamefont {K.~T.}\ \bibnamefont
  {Inoue}}, \bibinfo {author} {\bibfnamefont {R.}~\bibnamefont {Takahashi}},
  \bibinfo {author} {\bibfnamefont {T.}~\bibnamefont {Takahashi}}, \ and\
  \bibinfo {author} {\bibfnamefont {T.}~\bibnamefont {Ishiyama}},\ }\href
  {\doibase 10.1093/mnras/stv194} {\bibfield  {journal} {\bibinfo  {journal}
  {Mon. Not. Roy. Astron. Soc.}\ }\textbf {\bibinfo {volume} {448}},\ \bibinfo
  {pages} {2704} (\bibinfo {year} {2015})},\ \Eprint
  {http://arxiv.org/abs/1409.1326} {arXiv:1409.1326 [astro-ph.CO]} \BibitemShut
  {NoStop}%
\bibitem [{\citenamefont {Kuijken}\ \emph {et~al.}(2015)\citenamefont {Kuijken}
  \emph {et~al.}}]{Kuijken:2015vca}%
  \BibitemOpen
  \bibfield  {author} {\bibinfo {author} {\bibfnamefont {K.}~\bibnamefont
  {Kuijken}} \emph {et~al.},\ }\href {\doibase 10.1093/mnras/stv2140}
  {\bibfield  {journal} {\bibinfo  {journal} {Mon. Not. Roy. Astron. Soc.}\
  }\textbf {\bibinfo {volume} {454}},\ \bibinfo {pages} {3500} (\bibinfo {year}
  {2015})},\ \Eprint {http://arxiv.org/abs/1507.00738} {arXiv:1507.00738
  [astro-ph.CO]} \BibitemShut {NoStop}%
\bibitem [{\citenamefont {Conti}\ \emph {et~al.}(2016)\citenamefont {Conti},
  \citenamefont {Herbonnet}, \citenamefont {Hoekstra}, \citenamefont {Merten},
  \citenamefont {Miller},\ and\ \citenamefont {Viola}}]{Conti:2016gav}%
  \BibitemOpen
  \bibfield  {author} {\bibinfo {author} {\bibfnamefont {I.~F.}\ \bibnamefont
  {Conti}}, \bibinfo {author} {\bibfnamefont {R.}~\bibnamefont {Herbonnet}},
  \bibinfo {author} {\bibfnamefont {H.}~\bibnamefont {Hoekstra}}, \bibinfo
  {author} {\bibfnamefont {J.}~\bibnamefont {Merten}}, \bibinfo {author}
  {\bibfnamefont {L.}~\bibnamefont {Miller}}, \ and\ \bibinfo {author}
  {\bibfnamefont {M.}~\bibnamefont {Viola}},\ }\href@noop {} {\  (\bibinfo
  {year} {2016})},\ \Eprint {http://arxiv.org/abs/1606.05337} {arXiv:1606.05337
  [astro-ph.CO]} \BibitemShut {NoStop}%
\bibitem [{\citenamefont {Kitching}\ \emph
  {et~al.}(2016{\natexlab{b}})\citenamefont {Kitching}, \citenamefont {Verde},
  \citenamefont {Heavens},\ and\ \citenamefont {Jimenez}}]{Kitching:2016hvn}%
  \BibitemOpen
  \bibfield  {author} {\bibinfo {author} {\bibfnamefont {T.~D.}\ \bibnamefont
  {Kitching}}, \bibinfo {author} {\bibfnamefont {L.}~\bibnamefont {Verde}},
  \bibinfo {author} {\bibfnamefont {A.~F.}\ \bibnamefont {Heavens}}, \ and\
  \bibinfo {author} {\bibfnamefont {R.}~\bibnamefont {Jimenez}},\ }\href
  {\doibase 10.1093/mnras/stw707} {\bibfield  {journal} {\bibinfo  {journal}
  {Mon. Not. Roy. Astron. Soc.}\ }\textbf {\bibinfo {volume} {459}},\ \bibinfo
  {pages} {971} (\bibinfo {year} {2016}{\natexlab{b}})},\ \Eprint
  {http://arxiv.org/abs/1602.02960} {arXiv:1602.02960 [astro-ph.CO]}
  \BibitemShut {NoStop}%
\bibitem [{\citenamefont {Joudaki}\ \emph
  {et~al.}(2016{\natexlab{b}})\citenamefont {Joudaki} \emph
  {et~al.}}]{Joudaki:2016mvz}%
  \BibitemOpen
  \bibfield  {author} {\bibinfo {author} {\bibfnamefont {S.}~\bibnamefont
  {Joudaki}} \emph {et~al.},\ }\href@noop {} {\  (\bibinfo {year}
  {2016}{\natexlab{b}})},\ \Eprint {http://arxiv.org/abs/1601.05786}
  {arXiv:1601.05786 [astro-ph.CO]} \BibitemShut {NoStop}%
\bibitem [{\citenamefont {Schneider}(2016)}]{Schneider:2016uqi}%
  \BibitemOpen
  \bibfield  {author} {\bibinfo {author} {\bibfnamefont {A.}~\bibnamefont
  {Schneider}},\ }\href {\doibase 10.1088/1475-7516/2016/04/059} {\bibfield
  {journal} {\bibinfo  {journal} {JCAP}\ }\textbf {\bibinfo {volume} {1604}},\
  \bibinfo {pages} {059} (\bibinfo {year} {2016})},\ \Eprint
  {http://arxiv.org/abs/1601.07553} {arXiv:1601.07553 [astro-ph.CO]}
  \BibitemShut {NoStop}%
\bibitem [{\citenamefont {Polisensky}\ and\ \citenamefont
  {Ricotti}(2011)}]{Polisensky:2010rw}%
  \BibitemOpen
  \bibfield  {author} {\bibinfo {author} {\bibfnamefont {E.}~\bibnamefont
  {Polisensky}}\ and\ \bibinfo {author} {\bibfnamefont {M.}~\bibnamefont
  {Ricotti}},\ }\href {\doibase 10.1103/PhysRevD.83.043506} {\bibfield
  {journal} {\bibinfo  {journal} {Phys. Rev.}\ }\textbf {\bibinfo {volume}
  {D83}},\ \bibinfo {pages} {043506} (\bibinfo {year} {2011})},\ \Eprint
  {http://arxiv.org/abs/1004.1459} {arXiv:1004.1459 [astro-ph.CO]} \BibitemShut
  {NoStop}%
\bibitem [{\citenamefont {Wolf}\ \emph {et~al.}(2010)\citenamefont {Wolf},
  \citenamefont {Martinez}, \citenamefont {Bullock}, \citenamefont
  {Kaplinghat}, \citenamefont {Geha}, \citenamefont {Mu{\~{n}}oz},
  \citenamefont {Simon},\ and\ \citenamefont {Avedo}}]{Wolf:2009a}%
  \BibitemOpen
  \bibfield  {author} {\bibinfo {author} {\bibfnamefont {J.}~\bibnamefont
  {Wolf}}, \bibinfo {author} {\bibfnamefont {G.~D.}\ \bibnamefont {Martinez}},
  \bibinfo {author} {\bibfnamefont {J.~S.}\ \bibnamefont {Bullock}}, \bibinfo
  {author} {\bibfnamefont {M.}~\bibnamefont {Kaplinghat}}, \bibinfo {author}
  {\bibfnamefont {M.}~\bibnamefont {Geha}}, \bibinfo {author} {\bibfnamefont
  {R.~R.}\ \bibnamefont {Mu{\~{n}}oz}}, \bibinfo {author} {\bibfnamefont
  {J.~D.}\ \bibnamefont {Simon}}, \ and\ \bibinfo {author} {\bibfnamefont
  {F.~F.}\ \bibnamefont {Avedo}},\ }\href {\doibase
  10.1111/j.1365-2966.2010.16753.x} {\bibfield  {journal} {\bibinfo  {journal}
  {Monthly Notices of the Royal Astronomical Society}\ ,\ \bibinfo {pages}
  {no}} (\bibinfo {year} {2010})}\BibitemShut {NoStop}%
\bibitem [{\citenamefont {Bechtol}\ \emph {et~al.}(2015)\citenamefont {Bechtol}
  \emph {et~al.}}]{Bechtol:2015cbp}%
  \BibitemOpen
  \bibfield  {author} {\bibinfo {author} {\bibfnamefont {K.}~\bibnamefont
  {Bechtol}} \emph {et~al.} (\bibinfo {collaboration} {DES}),\ }\href {\doibase
  10.1088/0004-637X/807/1/50} {\bibfield  {journal} {\bibinfo  {journal}
  {Astrophys. J.}\ }\textbf {\bibinfo {volume} {807}},\ \bibinfo {pages} {50}
  (\bibinfo {year} {2015})},\ \Eprint {http://arxiv.org/abs/1503.02584}
  {arXiv:1503.02584 [astro-ph.GA]} \BibitemShut {NoStop}%
\bibitem [{\citenamefont {Drlica-Wagner}\ \emph {et~al.}(2015)\citenamefont
  {Drlica-Wagner} \emph {et~al.}}]{Drlica-Wagner:2015ufc}%
  \BibitemOpen
  \bibfield  {author} {\bibinfo {author} {\bibfnamefont {A.}~\bibnamefont
  {Drlica-Wagner}} \emph {et~al.} (\bibinfo {collaboration} {DES}),\ }\href
  {\doibase 10.1088/0004-637X/813/2/109} {\bibfield  {journal} {\bibinfo
  {journal} {Astrophys. J.}\ }\textbf {\bibinfo {volume} {813}},\ \bibinfo
  {pages} {109} (\bibinfo {year} {2015})},\ \Eprint
  {http://arxiv.org/abs/1508.03622} {arXiv:1508.03622 [astro-ph.GA]}
  \BibitemShut {NoStop}%
\bibitem [{\citenamefont {Boylan-Kolchin}\ \emph {et~al.}(2010)\citenamefont
  {Boylan-Kolchin}, \citenamefont {Springel}, \citenamefont {White},\ and\
  \citenamefont {Jenkins}}]{BoylanKolchin:2009an}%
  \BibitemOpen
  \bibfield  {author} {\bibinfo {author} {\bibfnamefont {M.}~\bibnamefont
  {Boylan-Kolchin}}, \bibinfo {author} {\bibfnamefont {V.}~\bibnamefont
  {Springel}}, \bibinfo {author} {\bibfnamefont {S.~D.~M.}\ \bibnamefont
  {White}}, \ and\ \bibinfo {author} {\bibfnamefont {A.}~\bibnamefont
  {Jenkins}},\ }\href {\doibase 10.1111/j.1365-2966.2010.16774.x} {\bibfield
  {journal} {\bibinfo  {journal} {Mon. Not. Roy. Astron. Soc.}\ }\textbf
  {\bibinfo {volume} {406}},\ \bibinfo {pages} {896} (\bibinfo {year}
  {2010})},\ \Eprint {http://arxiv.org/abs/0911.4484} {arXiv:0911.4484
  [astro-ph.CO]} \BibitemShut {NoStop}%
\bibitem [{\citenamefont {Tollerud}\ \emph {et~al.}(2008)\citenamefont
  {Tollerud}, \citenamefont {Bullock}, \citenamefont {Strigari},\ and\
  \citenamefont {Willman}}]{Tollerud:2008ze}%
  \BibitemOpen
  \bibfield  {author} {\bibinfo {author} {\bibfnamefont {E.~J.}\ \bibnamefont
  {Tollerud}}, \bibinfo {author} {\bibfnamefont {J.~S.}\ \bibnamefont
  {Bullock}}, \bibinfo {author} {\bibfnamefont {L.~E.}\ \bibnamefont
  {Strigari}}, \ and\ \bibinfo {author} {\bibfnamefont {B.}~\bibnamefont
  {Willman}},\ }\href {\doibase 10.1086/592102} {\bibfield  {journal} {\bibinfo
   {journal} {Astrophys. J.}\ }\textbf {\bibinfo {volume} {688}},\ \bibinfo
  {pages} {277} (\bibinfo {year} {2008})},\ \Eprint
  {http://arxiv.org/abs/0806.4381} {arXiv:0806.4381 [astro-ph]} \BibitemShut
  {NoStop}%
\bibitem [{\citenamefont {Brooks}\ and\ \citenamefont
  {Zolotov}(2014)}]{Brooks:2012a}%
  \BibitemOpen
  \bibfield  {author} {\bibinfo {author} {\bibfnamefont {A.~M.}\ \bibnamefont
  {Brooks}}\ and\ \bibinfo {author} {\bibfnamefont {A.}~\bibnamefont
  {Zolotov}},\ }\href {\doibase 10.1088/0004-637x/786/2/87} {\bibfield
  {journal} {\bibinfo  {journal} {The Astrophysical Journal}\ }\textbf
  {\bibinfo {volume} {786}},\ \bibinfo {pages} {87} (\bibinfo {year}
  {2014})}\BibitemShut {NoStop}%
\bibitem [{\citenamefont {Schneider}(2015)}]{Schneider:2014rda}%
  \BibitemOpen
  \bibfield  {author} {\bibinfo {author} {\bibfnamefont {A.}~\bibnamefont
  {Schneider}},\ }\href {\doibase 10.1093/mnras/stv1169} {\bibfield  {journal}
  {\bibinfo  {journal} {Mon. Not. Roy. Astron. Soc.}\ }\textbf {\bibinfo
  {volume} {451}},\ \bibinfo {pages} {3117} (\bibinfo {year} {2015})},\ \Eprint
  {http://arxiv.org/abs/1412.2133} {arXiv:1412.2133 [astro-ph.CO]} \BibitemShut
  {NoStop}%
\bibitem [{\citenamefont {Guo}\ \emph {et~al.}(2010)\citenamefont {Guo},
  \citenamefont {White}, \citenamefont {Li},\ and\ \citenamefont
  {Boylan-Kolchin}}]{Guo:2009a}%
  \BibitemOpen
  \bibfield  {author} {\bibinfo {author} {\bibfnamefont {Q.}~\bibnamefont
  {Guo}}, \bibinfo {author} {\bibfnamefont {S.}~\bibnamefont {White}}, \bibinfo
  {author} {\bibfnamefont {C.}~\bibnamefont {Li}}, \ and\ \bibinfo {author}
  {\bibfnamefont {M.}~\bibnamefont {Boylan-Kolchin}},\ }\href {\doibase
  10.1111/j.1365-2966.2010.16341.x} {\bibfield  {journal} {\bibinfo  {journal}
  {Monthly Notices of the Royal Astronomical Society}\ } (\bibinfo {year}
  {2010}),\ 10.1111/j.1365-2966.2010.16341.x}\BibitemShut {NoStop}%
\bibitem [{\citenamefont {Springel}\ \emph {et~al.}(2008)\citenamefont
  {Springel}, \citenamefont {Wang}, \citenamefont {Vogelsberger}, \citenamefont
  {Ludlow}, \citenamefont {Jenkins}, \citenamefont {Helmi}, \citenamefont
  {Navarro}, \citenamefont {Frenk},\ and\ \citenamefont
  {White}}]{Springel:2008cc}%
  \BibitemOpen
  \bibfield  {author} {\bibinfo {author} {\bibfnamefont {V.}~\bibnamefont
  {Springel}}, \bibinfo {author} {\bibfnamefont {J.}~\bibnamefont {Wang}},
  \bibinfo {author} {\bibfnamefont {M.}~\bibnamefont {Vogelsberger}}, \bibinfo
  {author} {\bibfnamefont {A.}~\bibnamefont {Ludlow}}, \bibinfo {author}
  {\bibfnamefont {A.}~\bibnamefont {Jenkins}}, \bibinfo {author} {\bibfnamefont
  {A.}~\bibnamefont {Helmi}}, \bibinfo {author} {\bibfnamefont {J.~F.}\
  \bibnamefont {Navarro}}, \bibinfo {author} {\bibfnamefont {C.~S.}\
  \bibnamefont {Frenk}}, \ and\ \bibinfo {author} {\bibfnamefont {S.~D.~M.}\
  \bibnamefont {White}},\ }\href {\doibase 10.1111/j.1365-2966.2008.14066.x}
  {\bibfield  {journal} {\bibinfo  {journal} {Mon. Not. Roy. Astron. Soc.}\
  }\textbf {\bibinfo {volume} {391}},\ \bibinfo {pages} {1685} (\bibinfo {year}
  {2008})},\ \Eprint {http://arxiv.org/abs/0809.0898} {arXiv:0809.0898
  [astro-ph]} \BibitemShut {NoStop}%
\bibitem [{\citenamefont {Press}\ and\ \citenamefont
  {Schechter}(1974)}]{Press:1973iz}%
  \BibitemOpen
  \bibfield  {author} {\bibinfo {author} {\bibfnamefont {W.~H.}\ \bibnamefont
  {Press}}\ and\ \bibinfo {author} {\bibfnamefont {P.}~\bibnamefont
  {Schechter}},\ }\href {\doibase 10.1086/152650} {\bibfield  {journal}
  {\bibinfo  {journal} {Astrophys. J.}\ }\textbf {\bibinfo {volume} {187}},\
  \bibinfo {pages} {425} (\bibinfo {year} {1974})}\BibitemShut {NoStop}%
\bibitem [{\citenamefont {Mangano}\ \emph {et~al.}(2005)\citenamefont
  {Mangano}, \citenamefont {Miele}, \citenamefont {Pastor}, \citenamefont
  {Pinto}, \citenamefont {Pisanti},\ and\ \citenamefont
  {Serpico}}]{Mangano:2005cc}%
  \BibitemOpen
  \bibfield  {author} {\bibinfo {author} {\bibfnamefont {G.}~\bibnamefont
  {Mangano}}, \bibinfo {author} {\bibfnamefont {G.}~\bibnamefont {Miele}},
  \bibinfo {author} {\bibfnamefont {S.}~\bibnamefont {Pastor}}, \bibinfo
  {author} {\bibfnamefont {T.}~\bibnamefont {Pinto}}, \bibinfo {author}
  {\bibfnamefont {O.}~\bibnamefont {Pisanti}}, \ and\ \bibinfo {author}
  {\bibfnamefont {P.~D.}\ \bibnamefont {Serpico}},\ }\href {\doibase
  10.1016/j.nuclphysb.2005.09.041} {\bibfield  {journal} {\bibinfo  {journal}
  {Nucl. Phys.}\ }\textbf {\bibinfo {volume} {B729}},\ \bibinfo {pages} {221}
  (\bibinfo {year} {2005})},\ \Eprint {http://arxiv.org/abs/hep-ph/0506164}
  {arXiv:hep-ph/0506164 [hep-ph]} \BibitemShut {NoStop}%
\bibitem [{\citenamefont {de~Salas}\ and\ \citenamefont
  {Pastor}(2016)}]{deSalas:2016ztq}%
  \BibitemOpen
  \bibfield  {author} {\bibinfo {author} {\bibfnamefont {P.~F.}\ \bibnamefont
  {de~Salas}}\ and\ \bibinfo {author} {\bibfnamefont {S.}~\bibnamefont
  {Pastor}},\ }\href {\doibase 10.1088/1475-7516/2016/07/051} {\bibfield
  {journal} {\bibinfo  {journal} {JCAP}\ }\textbf {\bibinfo {volume} {1607}},\
  \bibinfo {pages} {051} (\bibinfo {year} {2016})},\ \Eprint
  {http://arxiv.org/abs/1606.06986} {arXiv:1606.06986 [hep-ph]} \BibitemShut
  {NoStop}%
\bibitem [{\citenamefont {Audren}\ \emph {et~al.}(2013)\citenamefont {Audren},
  \citenamefont {Lesgourgues}, \citenamefont {Benabed},\ and\ \citenamefont
  {Prunet}}]{Audren:2012wb}%
  \BibitemOpen
  \bibfield  {author} {\bibinfo {author} {\bibfnamefont {B.}~\bibnamefont
  {Audren}}, \bibinfo {author} {\bibfnamefont {J.}~\bibnamefont {Lesgourgues}},
  \bibinfo {author} {\bibfnamefont {K.}~\bibnamefont {Benabed}}, \ and\
  \bibinfo {author} {\bibfnamefont {S.}~\bibnamefont {Prunet}},\ }\href
  {\doibase 10.1088/1475-7516/2013/02/001} {\bibfield  {journal} {\bibinfo
  {journal} {JCAP}\ }\textbf {\bibinfo {volume} {1302}},\ \bibinfo {pages}
  {001} (\bibinfo {year} {2013})},\ \Eprint {http://arxiv.org/abs/1210.7183}
  {arXiv:1210.7183 [astro-ph.CO]} \BibitemShut {NoStop}%
\bibitem [{\citenamefont {Lesgourgues}(2011)}]{Lesgourgues:2011re}%
  \BibitemOpen
  \bibfield  {author} {\bibinfo {author} {\bibfnamefont {J.}~\bibnamefont
  {Lesgourgues}},\ }\href@noop {} {\  (\bibinfo {year} {2011})},\ \Eprint
  {http://arxiv.org/abs/1104.2932} {arXiv:1104.2932 [astro-ph.IM]} \BibitemShut
  {NoStop}%
\bibitem [{\citenamefont {Lewis}\ \emph {et~al.}(2000)\citenamefont {Lewis},
  \citenamefont {Challinor},\ and\ \citenamefont {Lasenby}}]{Lewis:1999bs}%
  \BibitemOpen
  \bibfield  {author} {\bibinfo {author} {\bibfnamefont {A.}~\bibnamefont
  {Lewis}}, \bibinfo {author} {\bibfnamefont {A.}~\bibnamefont {Challinor}}, \
  and\ \bibinfo {author} {\bibfnamefont {A.}~\bibnamefont {Lasenby}},\ }\href
  {\doibase 10.1086/309179} {\bibfield  {journal} {\bibinfo  {journal}
  {Astrophys. J.}\ }\textbf {\bibinfo {volume} {538}},\ \bibinfo {pages} {473}
  (\bibinfo {year} {2000})},\ \Eprint {http://arxiv.org/abs/astro-ph/9911177}
  {arXiv:astro-ph/9911177 [astro-ph]} \BibitemShut {NoStop}%
\bibitem [{\citenamefont {Lewis}\ and\ \citenamefont
  {Bridle}(2002)}]{Lewis:2002ah}%
  \BibitemOpen
  \bibfield  {author} {\bibinfo {author} {\bibfnamefont {A.}~\bibnamefont
  {Lewis}}\ and\ \bibinfo {author} {\bibfnamefont {S.}~\bibnamefont {Bridle}},\
  }\href {\doibase 10.1103/PhysRevD.66.103511} {\bibfield  {journal} {\bibinfo
  {journal} {Phys. Rev.}\ }\textbf {\bibinfo {volume} {D66}},\ \bibinfo {pages}
  {103511} (\bibinfo {year} {2002})},\ \Eprint
  {http://arxiv.org/abs/astro-ph/0205436} {arXiv:astro-ph/0205436 [astro-ph]}
  \BibitemShut {NoStop}%
\bibitem [{\citenamefont {Moore}\ \emph
  {et~al.}(1999{\natexlab{b}})\citenamefont {Moore}, \citenamefont {Quinn},
  \citenamefont {Governato}, \citenamefont {Stadel},\ and\ \citenamefont
  {Lake}}]{Moore:1999gc}%
  \BibitemOpen
  \bibfield  {author} {\bibinfo {author} {\bibfnamefont {B.}~\bibnamefont
  {Moore}}, \bibinfo {author} {\bibfnamefont {T.~R.}\ \bibnamefont {Quinn}},
  \bibinfo {author} {\bibfnamefont {F.}~\bibnamefont {Governato}}, \bibinfo
  {author} {\bibfnamefont {J.}~\bibnamefont {Stadel}}, \ and\ \bibinfo {author}
  {\bibfnamefont {G.}~\bibnamefont {Lake}},\ }\href {\doibase
  10.1046/j.1365-8711.1999.03039.x} {\bibfield  {journal} {\bibinfo  {journal}
  {Mon. Not. Roy. Astron. Soc.}\ }\textbf {\bibinfo {volume} {310}},\ \bibinfo
  {pages} {1147} (\bibinfo {year} {1999}{\natexlab{b}})},\ \Eprint
  {http://arxiv.org/abs/astro-ph/9903164} {arXiv:astro-ph/9903164 [astro-ph]}
  \BibitemShut {NoStop}%
\bibitem [{\citenamefont {Bode}\ \emph {et~al.}(2001)\citenamefont {Bode},
  \citenamefont {Ostriker},\ and\ \citenamefont {Turok}}]{Bode:2000gq}%
  \BibitemOpen
  \bibfield  {author} {\bibinfo {author} {\bibfnamefont {P.}~\bibnamefont
  {Bode}}, \bibinfo {author} {\bibfnamefont {J.~P.}\ \bibnamefont {Ostriker}},
  \ and\ \bibinfo {author} {\bibfnamefont {N.}~\bibnamefont {Turok}},\ }\href
  {\doibase 10.1086/321541} {\bibfield  {journal} {\bibinfo  {journal}
  {Astrophys. J.}\ }\textbf {\bibinfo {volume} {556}},\ \bibinfo {pages} {93}
  (\bibinfo {year} {2001})},\ \Eprint {http://arxiv.org/abs/astro-ph/0010389}
  {arXiv:astro-ph/0010389 [astro-ph]} \BibitemShut {NoStop}%
\bibitem [{\citenamefont {Avila-Reese}\ \emph {et~al.}(2001)\citenamefont
  {Avila-Reese}, \citenamefont {Colin}, \citenamefont {Valenzuela},
  \citenamefont {D'Onghia},\ and\ \citenamefont {Firmani}}]{AvilaReese:2000hg}%
  \BibitemOpen
  \bibfield  {author} {\bibinfo {author} {\bibfnamefont {V.}~\bibnamefont
  {Avila-Reese}}, \bibinfo {author} {\bibfnamefont {P.}~\bibnamefont {Colin}},
  \bibinfo {author} {\bibfnamefont {O.}~\bibnamefont {Valenzuela}}, \bibinfo
  {author} {\bibfnamefont {E.}~\bibnamefont {D'Onghia}}, \ and\ \bibinfo
  {author} {\bibfnamefont {C.}~\bibnamefont {Firmani}},\ }\href {\doibase
  10.1086/322411} {\bibfield  {journal} {\bibinfo  {journal} {Astrophys. J.}\
  }\textbf {\bibinfo {volume} {559}},\ \bibinfo {pages} {516} (\bibinfo {year}
  {2001})},\ \Eprint {http://arxiv.org/abs/astro-ph/0010525}
  {arXiv:astro-ph/0010525 [astro-ph]} \BibitemShut {NoStop}%
\bibitem [{\citenamefont {Narayanan}\ \emph {et~al.}(2000)\citenamefont
  {Narayanan}, \citenamefont {Spergel}, \citenamefont {Dave},\ and\
  \citenamefont {Ma}}]{Narayanan:2000tp}%
  \BibitemOpen
  \bibfield  {author} {\bibinfo {author} {\bibfnamefont {V.~K.}\ \bibnamefont
  {Narayanan}}, \bibinfo {author} {\bibfnamefont {D.~N.}\ \bibnamefont
  {Spergel}}, \bibinfo {author} {\bibfnamefont {R.}~\bibnamefont {Dave}}, \
  and\ \bibinfo {author} {\bibfnamefont {C.-P.}\ \bibnamefont {Ma}},\ }\href
  {\doibase 10.1086/317269} {\bibfield  {journal} {\bibinfo  {journal}
  {Astrophys. J.}\ }\textbf {\bibinfo {volume} {543}},\ \bibinfo {pages} {L103}
  (\bibinfo {year} {2000})},\ \Eprint {http://arxiv.org/abs/astro-ph/0005095}
  {arXiv:astro-ph/0005095 [astro-ph]} \BibitemShut {NoStop}%
\bibitem [{\citenamefont {Viel}\ \emph {et~al.}(2005)\citenamefont {Viel},
  \citenamefont {Lesgourgues}, \citenamefont {Haehnelt}, \citenamefont
  {Matarrese},\ and\ \citenamefont {Riotto}}]{Viel:2005qj}%
  \BibitemOpen
  \bibfield  {author} {\bibinfo {author} {\bibfnamefont {M.}~\bibnamefont
  {Viel}}, \bibinfo {author} {\bibfnamefont {J.}~\bibnamefont {Lesgourgues}},
  \bibinfo {author} {\bibfnamefont {M.~G.}\ \bibnamefont {Haehnelt}}, \bibinfo
  {author} {\bibfnamefont {S.}~\bibnamefont {Matarrese}}, \ and\ \bibinfo
  {author} {\bibfnamefont {A.}~\bibnamefont {Riotto}},\ }\href {\doibase
  10.1103/PhysRevD.71.063534} {\bibfield  {journal} {\bibinfo  {journal} {Phys.
  Rev.}\ }\textbf {\bibinfo {volume} {D71}},\ \bibinfo {pages} {063534}
  (\bibinfo {year} {2005})},\ \Eprint {http://arxiv.org/abs/astro-ph/0501562}
  {arXiv:astro-ph/0501562 [astro-ph]} \BibitemShut {NoStop}%
\bibitem [{\citenamefont {Viel}\ \emph {et~al.}(2013)\citenamefont {Viel},
  \citenamefont {Becker}, \citenamefont {Bolton},\ and\ \citenamefont
  {Haehnelt}}]{Viel:2013apy}%
  \BibitemOpen
  \bibfield  {author} {\bibinfo {author} {\bibfnamefont {M.}~\bibnamefont
  {Viel}}, \bibinfo {author} {\bibfnamefont {G.~D.}\ \bibnamefont {Becker}},
  \bibinfo {author} {\bibfnamefont {J.~S.}\ \bibnamefont {Bolton}}, \ and\
  \bibinfo {author} {\bibfnamefont {M.~G.}\ \bibnamefont {Haehnelt}},\ }\href
  {\doibase 10.1103/PhysRevD.88.043502} {\bibfield  {journal} {\bibinfo
  {journal} {Phys. Rev.}\ }\textbf {\bibinfo {volume} {D88}},\ \bibinfo {pages}
  {043502} (\bibinfo {year} {2013})},\ \Eprint {http://arxiv.org/abs/1306.2314}
  {arXiv:1306.2314 [astro-ph.CO]} \BibitemShut {NoStop}%
\bibitem [{\citenamefont {Baur}\ \emph {et~al.}(2016)\citenamefont {Baur},
  \citenamefont {Palanque-Delabrouille}, \citenamefont {Yeche}, \citenamefont
  {Magneville},\ and\ \citenamefont {Viel}}]{Baur:2015jsy}%
  \BibitemOpen
  \bibfield  {author} {\bibinfo {author} {\bibfnamefont {J.}~\bibnamefont
  {Baur}}, \bibinfo {author} {\bibfnamefont {N.}~\bibnamefont
  {Palanque-Delabrouille}}, \bibinfo {author} {\bibfnamefont {C.}~\bibnamefont
  {Yeche}}, \bibinfo {author} {\bibfnamefont {C.}~\bibnamefont {Magneville}}, \
  and\ \bibinfo {author} {\bibfnamefont {M.}~\bibnamefont {Viel}},\ }\bibfield
  {booktitle} {\emph {\bibinfo {booktitle} {SDSS-IV Collaboration Meeting, July
  20-23, 2015}},\ }\href {\doibase 10.1088/1475-7516/2016/08/012} {\bibfield
  {journal} {\bibinfo  {journal} {JCAP}\ }\textbf {\bibinfo {volume} {1608}},\
  \bibinfo {pages} {012} (\bibinfo {year} {2016})},\ \Eprint
  {http://arxiv.org/abs/1512.01981} {arXiv:1512.01981 [astro-ph.CO]}
  \BibitemShut {NoStop}%
\bibitem [{\citenamefont {Irsic}\ \emph {et~al.}(2017)\citenamefont {Irsic}
  \emph {et~al.}}]{Irsic:2017ixq}%
  \BibitemOpen
  \bibfield  {author} {\bibinfo {author} {\bibfnamefont {V.}~\bibnamefont
  {Irsic}} \emph {et~al.},\ }\href@noop {} {\  (\bibinfo {year} {2017})},\
  \Eprint {http://arxiv.org/abs/1702.01764} {arXiv:1702.01764 [astro-ph.CO]}
  \BibitemShut {NoStop}%
\bibitem [{\citenamefont {Yeche}\ \emph {et~al.}(2017)\citenamefont {Yeche},
  \citenamefont {Palanque-Delabrouille}, \citenamefont {Baur},\ and\
  \citenamefont {BourBoux}}]{Yeche:2017upn}%
  \BibitemOpen
  \bibfield  {author} {\bibinfo {author} {\bibfnamefont {C.}~\bibnamefont
  {Yeche}}, \bibinfo {author} {\bibfnamefont {N.}~\bibnamefont
  {Palanque-Delabrouille}}, \bibinfo {author} {\bibfnamefont {J.~.}\
  \bibnamefont {Baur}}, \ and\ \bibinfo {author} {\bibfnamefont {H.~d. M.~d.}\
  \bibnamefont {BourBoux}},\ }\href@noop {} {\  (\bibinfo {year} {2017})},\
  \Eprint {http://arxiv.org/abs/1702.03314} {arXiv:1702.03314 [astro-ph.CO]}
  \BibitemShut {NoStop}%
\bibitem [{\citenamefont {Lopez-Honorez}\ \emph {et~al.}(2017)\citenamefont
  {Lopez-Honorez}, \citenamefont {Mena}, \citenamefont {Palomares-Ruiz},\ and\
  \citenamefont {Domingo}}]{Lopez-Honorez:2017csg}%
  \BibitemOpen
  \bibfield  {author} {\bibinfo {author} {\bibfnamefont {L.}~\bibnamefont
  {Lopez-Honorez}}, \bibinfo {author} {\bibfnamefont {O.}~\bibnamefont {Mena}},
  \bibinfo {author} {\bibfnamefont {S.}~\bibnamefont {Palomares-Ruiz}}, \ and\
  \bibinfo {author} {\bibfnamefont {P.~V.}\ \bibnamefont {Domingo}},\
  }\href@noop {} {\  (\bibinfo {year} {2017})},\ \Eprint
  {http://arxiv.org/abs/1703.02302} {arXiv:1703.02302 [astro-ph.CO]}
  \BibitemShut {NoStop}%
\bibitem [{\citenamefont {Efstathiou}(2014)}]{Efstathiou:2013via}%
  \BibitemOpen
  \bibfield  {author} {\bibinfo {author} {\bibfnamefont {G.}~\bibnamefont
  {Efstathiou}},\ }\href {\doibase 10.1093/mnras/stu278} {\bibfield  {journal}
  {\bibinfo  {journal} {Mon. Not. Roy. Astron. Soc.}\ }\textbf {\bibinfo
  {volume} {440}},\ \bibinfo {pages} {1138} (\bibinfo {year} {2014})},\ \Eprint
  {http://arxiv.org/abs/1311.3461} {arXiv:1311.3461 [astro-ph.CO]} \BibitemShut
  {NoStop}%
\bibitem [{\citenamefont {Riess}\ \emph {et~al.}(2016)\citenamefont {Riess}
  \emph {et~al.}}]{Riess:2016jrr}%
  \BibitemOpen
  \bibfield  {author} {\bibinfo {author} {\bibfnamefont {A.~G.}\ \bibnamefont
  {Riess}} \emph {et~al.},\ }\href {\doibase 10.3847/0004-637X/826/1/56}
  {\bibfield  {journal} {\bibinfo  {journal} {Astrophys. J.}\ }\textbf
  {\bibinfo {volume} {826}},\ \bibinfo {pages} {56} (\bibinfo {year} {2016})},\
  \Eprint {http://arxiv.org/abs/1604.01424} {arXiv:1604.01424 [astro-ph.CO]}
  \BibitemShut {NoStop}%
\end{thebibliography}

\end{document}